\documentclass[showpacs,preprintnumbers,amsmath,amssymb,nofootinbib]{revtex4-1}
\usepackage{graphicx}
\usepackage{bbm}
\usepackage[T1]{fontenc}
\usepackage[latin1]{inputenc}
\usepackage{bezier, epsfig}
\usepackage{mathrsfs}
\newcommand{\Vek}[1]{\mbox{\boldmath$#1$\unboldmath}}

\newcommand{\bi}{\bigskip}
\newcommand{\no}{\noindent}
\newcommand{\be}{\begin{equation}}
 \newcommand{\ee}{\end{equation}}
 \newcommand{\bea}{\begin{eqnarray}}
 \newcommand{\eea}{\end{eqnarray}}

\newcommand{\hk}{\hspace{0.1cm}}

\newcommand{\rk}{\right)}
\newcommand{\lk}{\left(}

\newcommand{\sli}{\sum\limits}

\newcommand{\il}{\int\limits}

\newcommand{\vA}{\vek{A}}
\newcommand{\vx}{\vek{x}}

\newcommand{\vR}{\vek{R}}

\newcommand{\vy}{\vek{y}}

\newcommand{\vB}{\vek{B}}

\newcommand{\ve}{\vek{e}}

\newcommand{\vek}[1]{\mathbf{#1}}

\newcommand{\kvec}[1]{\mbox{\boldmath$\scriptstyle{#1}$\unboldmath}}

\newcommand{\kvx}{{\kvec{x}}}
\newcommand{\kvR}{{\kvec{R}}}

\newcommand{\tr}{\mathrm{tr}}

\begin{document}


\title{On the temporal Wilson loop in the Hamiltonian approach in Coulomb gauge}

\author{H.~Reinhardt}\email{hugo.reinhardt@uni-tuebingen.de}
\author{M.~Quandt}\email{markus.quandt@uni-tuebingen.de}
\author{G.Burgio}\email{giuseppe.burgio@uni-tuebingen.de}
\affiliation{%
Institut f\"ur Theoretische Physik\\
Auf der Morgenstelle 14\\
D-72076 T\"ubingen\\
Germany
}%

\date{\today}

\begin{abstract}
We investigate the temporal Wilson loop using the Hamiltonian approach to
Yang-Mills theory. In simple cases such as the Abelian theory  or 
the non-Abelian theory in $(1+1)$ dimensions, the known results can be 
reproduced using unitary transformations to take care of time evolution. 
We show how Coulomb gauge can be used for an alternative solution if the 
exact ground state wave functional is known. In the most interesting case 
of Yang-Mills theory in $(3+1)$ dimensions, the vacuum wave functional
is not known, but recent variational approaches in Coulomb gauge 
give a decent approximation. We use this formulation to compute the temporal 
Wilson loop and find that the Wilson and Coulomb string tension agree within 
our approximation scheme. Possible improvements of these findings are 
briefly discussed.  
\end{abstract}

\maketitle
\section{Introduction}
\label{intro}

\no
In recent years there has been a renewed interest in studying Yang-Mills
theory in Coulomb gauge, both in the continuum \cite{Zwanziger:2002sh} and on 
the lattice \cite{Langfeld:2004qs, Quandt:2007qd, Voigt:2007wd, Voigt:2008rr,%
Burgio:2008jr, Burgio:2009xp, Quandt:2010yq}. These developments are mainly based 
on the fact that the consequences of the Gribov-Zwanziger picture of 
confinement \cite{Gribov:1977wm, Zwanziger:1995cv} are 
particularly transparent in Coulomb gauge: Physical states in the 
Hamiltonian formulation have to obey Gauss' law, without the need of 
additional restrictions such as a vanishing color charge based, in turn, 
on the assumption of a globally conserved BRST charge. Furthermore, since Gauss' 
law can be resolved explicitly in Coulomb gauge, any renormalizable \emph{Ansatz} 
for a vacuum wave functional is admissible in this gauge without having to 
explicitly construct the physical Hilbert space \cite{Burgio:1999tg}, 
which is a key fact to make variational approaches viable. (The equivalent 
Gupta-Bleuler type of constraints  in covariant gauges have no such 
simple resolution.)

Consequently, much work was devoted to a variational solution of the 
Yang-Mills Schr\"odinger equation in Coulomb gauge
\cite{Szczepaniak:2001rg,Feuchter:2004mk,Feuchter:2004gb,Reinhardt:2004mm,%
Epple:2006hv,Epple:2007ut,Reinhardt:2008ij,Schleifenbaum:2006bq,%
Reinhardt:2007wh}.
Using Gaussian type of Ans\"atze for the vacuum wave functional, a set of
Dyson-Schwinger equations for the gluon and ghost propagators was derived
by minimizing the vacuum energy density. An infrared analysis
\cite{Schleifenbaum:2006bq}
of these equations exhibits solutions in accord with the Gribov-Zwanziger
confinement scenario. Imposing Zwanziger's horizon condition one finds an
infrared diverging gluon energy and a linear rising static quark (Coulomb)
potential \cite{Schleifenbaum:2006bq} and also a perimeter law for the 't Hooft
loop \cite{Reinhardt:2007wh}. These infrared properties are reproduced by a
full
numerical solution of the Dyson-Schwinger equations over the entire momentum
regime \cite{Epple:2006hv}, and are also supported by lattice calculations
\cite{Quandt:2007qd,Burgio:2008jr}. Moreover, since the inverse of the 
ghost dressing function can be identified with the dielectric function of the 
Yang-Mills vacuum \cite{Reinhardt:2008ek}, the latter behaves 
in the infra-red, by the horizon condition, like a perfect color 
dia-electric medium, i.e.~a dual superconductor. This suggests that the 
linearly rising Coulomb potential is a manifestation of confinement via the dual 
Meissner effect.

These nice physical conclusions are, however, a bit premature. Using 
variational arguments, one can show that the Coulomb potential is only 
an \emph{upper bound} for the true potential between heavy quarks, which 
implies "no confinement without Coulomb confinement" \cite{Zwanziger:2002sh}, 
but not the converse. 
To get a \emph{sufficient} criterion for confinement, one has to study the true
potential extracted from the temporal Wilson loop. 
In the non-Abelian continuum theory, the Wilson loop is difficult to 
calculate because of path ordering. In ref.~\cite{Pak:2009em}, the spatial 
Wilson loop was computed from a Dyson--Schwinger equation, which has 
previously been derived in ref.~\cite{Erickson:1999qv} in the context of 
supersymmetric Yang-Mills theory. Although a linearly rising potential 
was extracted from the obtained Wilson loop, it is not clear to which 
extend this approach also applies to non-supersymmetric Yang-Mills theory.

As was clarified in ref.~\cite{Haagensen:1997pi}, it is by no means trivial to 
extract the static potential from the Wilson loop even in simple cases such 
as QED, due to the presence of singular self-energy terms. Another 
attempt to extract the static Coulomb potential of QED from the Wilson loop
was undertaken in ref.~\cite{Gaete:1998vr}, where a partial wave expansion
has been used. We have not been able to reproduce the result of 
ref.~\cite{Gaete:1998vr} for the static potential in QED, and we 
reconsider the relevant partial wave expansion in appendix \ref{appA}. 
In the present paper, we will instead show that the proper and elegant way to 
extract the QED potential from the Wilson loop should be  based on a 
separation of longitudinal and transversal degrees of freedom. In Coulomb 
gauge, this separation is accomplished \emph{automatically}, which is the 
main reason why this particular gauge is so convenient for our purposes.

In the present paper, we will attempt to extract the static quark potential
from the Wilson loop within the Hamiltonian approach,
using gauge-invariant and Coulomb gauge
techniques in exactly solvable cases, and approximate results from variational
calculations for the most interesting case of non-Abelian Yang-Mills theory 
in $D=3+1$.

Our paper is organized as follows: In the next section, we demonstrate 
how a unitary transformation similar to the one that leads to the 
interaction picture in quantum mechanics gives rise to an induced electric 
field that allows to compute the Wilson potential exactly in special cases.
To prepare for the non-Abelian theory in $(3+1)$ dimensions, we also show how 
Coulomb gauge can be incorporated  in these exactly solvable models. 
In section 3, we first treat the case of an Abelian $U(1)$ theory 
reproducing the usual Coulomb potential from Maxwell's theory. Section 4 
treats Yang-Mills theory in $(1+1)$ dimensions on a cylindrical spacetime
manifold. Finally, section five employs the known Gaussian wave functional 
from variational approaches to compute, with additional approximations, 
the Wilson string tension, which happens to agree with the Coulomb string 
tension within the given approximation scheme. We conclude with a brief 
summary and an outlook on further improvements of our calculation.

\begin {figure}
\begin{center}
\includegraphics[width=5cm]{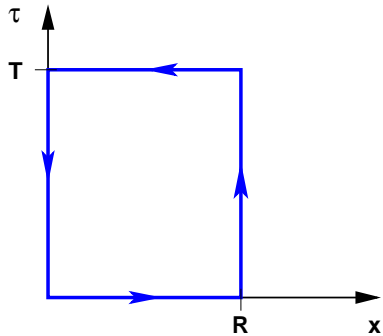}
\end{center}
\caption{Temporal rectangular Wilson loop.}
\label {fig1}
\end {figure}

\section{The temporal Wilson loop in the Hamilton approach}
\label{sec:invariant}
\no
Below, we investigate the temporal Wilson loop in the Hamiltonian approach to
$SU(N)$ Yang-Mills theory. This formulation assumes the Weyl gauge, $A_0 = 0$,
leaving a residual symmetry under spatial gauge rotations. Since the Wilson
loop operator is gauge invariant, it is not necessary to further fix this
residual symmetry in all cases. In fact, we will show that a gauge-invariant
Hamiltonian treatment of the temporal Wilson loop is successful in simple cases,
while more complex situations require gauge fixing as a technical tool to simplify
the separation of physical degrees of freedom.

\subsection{Physical interpretation}
\label{general}
We consider a planar time-like Wilson loop as shown in figure
\ref{fig1}. If the (Euclidean) time coordinate $\tau = x_0$ is unrestricted,
the Weyl gauge $A_0 = 0$ can be adopted and the Wilson loop operator
in the quark representation $r$ reads
\be
\label{G2}
W  =  \frac{1}{d_r} \,\tr_r \,\Big[ U[\vek{A}](\vek{0}, \vR\,;\, \tau = T ) \cdot
U[\vek{A}] (\vR, \vek{0}\,;\, \tau = 0 ) \Big]
=  \tr \Big[ U[\vek{A}] (\vR, \vek{0}\,;\, T)^\dagger \cdot
U[\vek{A}](\vR, \vek{0}\,;\, 0) \Big ] \hk ,
\ee
where $tr_r$ is the trace in the $SU(N)$ representation $r$, 
$d_r \equiv \tr_r\,\mathbbm{1}$ is its dimension and $\tr \equiv d_r^{-1}\,\tr$ 
denotes the normalised trace.
Furthermore, the parallel transporter along the path
$\mathscr{C}$ running from $\vx_1$ to $\vx_2$ at the time $x_0 = \tau$
is given by
\be
\label{G3}
U [\vek{A}] \lk \vx_2, \vx_1 \,;\, \tau \rk =\mathsf{P}\exp\left[ -
\il_{\mathscr{C}\,:\, \kvx_1 \to \kvx_2} d \vx \,\vA (\vx,\tau) \right]\,.
\ee
In this expression, we have absorbed the gauge coupling $g$ in the connection
$\vek{A}$, which entails that the Yang-Mills Hamiltonian (in the absence of
gauge fixing) takes the form
\begin{eqnarray}
\label{G2-X2}
H &=& \frac{g^2}{2} \int dx\,
\Vek{\Pi}^2 + \frac{1}{2 g^2} \int dx\, \mathbf{B}^2\,, \\[2mm]
\Pi^a_k(x) &=& - i \,\frac{\delta}{\delta A^a_k(x)}\,,\qquad\quad
B^a_k(x) = \epsilon_{k\ell m}\, \partial_\ell\, A^a_m(x) +
f^{abc}\, A_\ell^b(x)\,A_m^c(x)\,.
\end{eqnarray}
The time evolution of the gauge field in Euclidean space is then
\emph{formally}
\[
\vek{A}(\vek{x},\tau) = e^{- H \tau}\,\vek{A}(\vek{x},0)\,e^{H\tau}\,.
\]
and it is easy to see that the same semi-group law applies to the parallel
transporter, so that the \emph{vev} of the Wilson loop becomes
\be
\label{G8}
\langle \,W\, \rangle  = \tr\,
\langle\, 0 \,|\, U^\dagger(\vek{R})\,
e^{- (H - E_0) T} \,U(\vek{R}) \,|\, 0\, \rangle \hk
= \tr\,\langle\, \vek{R}\,|\,e^{- T (H-E_0)}\,|\,\vek{R}\,\rangle\,.
\ee
Here, $E_0$ is the energy of the YM vacuum $|\,0\,\rangle$ and the abbreviation
$U(\vek{R}) = U[\vek{A}](\vek{R},\vek{0}; \tau=0)$ for the straight line
transporter $\vek{0} \to \vek{R}$ made from gauge fields at time
$\tau=0$ was introduced.

\bigskip
The next step is to insert a complete set of energy eigenstates in the
matrix element eq.~(\ref{G8}). To do this properly, we have to recall that the
Hamiltonian eq.~(\ref{G2-X2}) is accompanied by a Hilbert space which decomposes
into charge subspaces by virtue of Gauss' law,
\begin{equation}
\hat{\Gamma}^a(\vek{z}) \,\Psi_{i_1,\ldots i_n}(\vek{A}\,;\,\vek{x}_1,\ldots, \vek{x}_n)
= - \left\{ \sum_{\ell=0}^n \sum_{\{k_\ell\}} \left[T^a_{(r_\ell)}\right]_{i_\ell k_\ell}\,
\delta(\vek{z}-\vek{x}_\ell)
\prod_{m \neq \ell}\delta_{i_m k_m}
\right\} \cdot
\,\Psi_{k_1,\ldots,k_n}(\vek{A}\,;\,\vek{x}_1,\ldots \vek{x}_n)
\label{gausslaw}
\end{equation}
where the sum is over the color spins of static quarks with color
representation $T_r$ located at positions $\vek{x}_1,\ldots,\vek{x}_n$.
Since the Gauss law operator $\Gamma^a(\vek{x}) \equiv \hat{\vek{D}}^{ab}(\vek{x})
\Vek{\Pi}^b(\vek{x})$ commutes with the (gauge-invariant) Hamiltonian,
the charge subspaces are invariant under the YM dynamics and mutually orthogonal.
By Schur's lemma, the Hamilton operator eq.~(\ref{G2-X2}) is thus proportional
to the color unit matrix when acting on any of the charge subspaces.
Moreover, the Gauss law operator generates time-independent gauge transformations
$U(\vek{x})$, i.e.~the iteration of eq.~(\ref{gausslaw}) states
(in the absence of a vacuum angle $\theta =0$)
that a physical state from the $n$-quark sector transforms under gauge rotations
as
\begin{equation}
\Psi_{i_1,\ldots,i_n}(\vek{A}^U\,;\,\vek{x}_1,\ldots,\vek{x}_n) =
\sum_{k_1,\ldots,k_n} \mathscr{D}^{(r_1)}_{i_1 k_1}[U(\vek{x}_1)]\cdots
\mathscr{D}^{(r_n)}_{i_n k_n}[U(\vek{x}_n)]\,\Psi_{k_1,\ldots,k_n}(\vek{A}\,;\,
\vek{x}_1,\ldots,\vek{x}_n)\,,
\label{gausslaw1}
\end{equation}
where $\mathscr{D}^{(r)}[\Omega]$ denotes the color matrix $\Omega$ in
representation $r$. For notational simplicity, we will henceforth assume that
the Wilson loop is in the fundamental representation, so that all representation
symbols $\mathscr{D}^{(r)}$ can be omitted.

It is now trivial to see that the \emph{Wilson state} 
\cite{Heinzl:2007kx,Heinzl:2008tv}
$|\vek{R}\rangle = U(\vek{R})\,|0\rangle$ from eq.~(\ref{G8}) corresponds to
the charge sector with a quark at position $\vek{R}$ and an antiquark in the
origin.
In fact, the identity
\begin{equation}
\Gamma^a(\vek{z}) U(\vek{R}) =
- \delta(\vek{z}-\vek{R})\,iT^a\,U(\vek{R}) + \delta(\vek{z}-\vek{0})\,
U(\vek{R})\,iT^a\,,
\label{charge0}
\end{equation}
implies the matrix relation
\begin{equation}
\Gamma^a(\vek{z})\,|\,\vek{R}\,\rangle_{ij} = -
\Big\{ \delta(\vek{z}-\vek{R})\,(iT^a)_{ik}\,\delta_{j\ell} - \delta(\vek{z})\,
\delta_{ik}\, (iT^a)_{\ell j}\Big\}\,|\,\vek{R}\,\rangle_{k\ell}\,,\label{charge1}
\end{equation}
which corresponds to eq.~(\ref{gausslaw}). For finite gauge transformations
$\Omega$ acting on the Wilson state, the equivalent of eq.~(\ref{gausslaw1}) is
\[
|\,\vek{R}\,\rangle_{ij} \to \Omega(\vek{R})_{ik}\cdot|\,\vek{R}\,
\rangle_{k\ell}\,\cdot \Omega(0)^{-1}_{\ell j}\,.
\]
Finally, for quark and antiquark in the same color representation, it is
often convenient to parallel transport the charges to the same
local color frame so that the charge density matrix acts on a single spin
index only. In the present case, this can be achieved by the identity
\begin{equation}
\big[U \cdot (iT^a)\big]_{i\ell} = \big[ U\,iT^a\,U^\dagger\big]_{ik}\cdot U_{k\ell}
\label{gaussid}
\end{equation}
so that eq.~(\ref{charge0}) takes the simpler form
\begin{equation}
\Gamma^a(\vek{z})\,U_{i\ell}(\vek{R}) =
\Big\{ \delta(\vek{z})\,[U(\vek{R})\,iT^a\,U(\vek{R})^\dagger]_{ik}
- \delta(\vek{z}-\vek{R})\,iT^a_{ik}\Big\}\,U_{k\ell}(\vek{R})
\equiv - {\rho}^a_{ik}(\vek{z})\cdot U_{k\ell}(\vek{R})\,,
\label{gauss2}
\end{equation}
where the second spin index ($\ell$) is not affected. As a consequence of the
parallel transport, the two charges at the endpoints of the Wilson line do
not exactly compensate
\be
\label{11-6}
Q^a = \int d\vek{z}\, \rho^a (\vek{z}) =
i\,\big[ \,U(\vek{R})\,T^a\,U(\vek{R})^\dagger - T^a\,\big] \neq 0
\ee
unless the color group is Abelian. (In any case, however, $\tr\,Q^a = 0$.)

\bigskip
Returning to eq.~(\ref{G8}), we insert a complete set of energy eigenstates
from all charge sectors, but only the states from the proper quark-antiquark
sector will give a non-vanishing contribution,
\be
\langle W \rangle  = \tr\, \sum_n\langle\, 0 \,|\, U^\dagger(\vek{R})\,
|\,n\,\rangle_{q\bar{q}}\cdot e^{- (H - E_0) T} \cdot {}_{q\bar{q}}\langle\,n\,|
\,U(\vek{R}) \,|\, 0\, \rangle =
\sum_{n}\,e^{- T (E_n^{(q\bar{q})} - E_0)}\,
\| \langle \vek{R}\,|\,n\,\rangle_{q\bar{q}}
\|^2\,,
\ee
where $\| \Omega\|^2 \equiv \tr(\Omega^\dagger\Omega)$. In the limit of large
Euclidean time extensions, we project out the static quark potential
\[
-\lim_{T \to \infty}\,T^{-1}\,\ln\,\langle W \rangle = E_0^{(q\bar{q})} - E_0
\equiv V(R)\,,
\]
provided that the Wilson line has \emph{non-vanishing overlap} with the true vacuum
in the proper quark-antiquark sector. (We will come back to this issue in
section \ref{sec:YM11} below.)

\subsection{Induced electric field}
To proceed, we use the spatial parallel transporter $U(\vek{R})$ to
induce a transformation similar to the one that leads to the interaction picture
in quantum mechanics. For instance, the new Hamilton operator becomes
\begin{equation}
\widetilde{H} = U^\dagger(\vek{R})\,H\,U(\vek{R}) \equiv \widetilde{H}(\vek{R})\,.
\label{T0}
\end{equation}
All transformed quantities will be $\vek{R}$-dependent and denoted, generally, 
by a tilde; they will also be color matrices in the representation chosen for 
the Wilson loop, although we will sometimes omit color indices for simplicity. 
For the transformed momentum operator, we obtain 
\begin{equation}
\widetilde{\Pi}^a_i(\vek{x}) = U^\dagger(\vek{R})\, \Pi^a_i(\vek{x})
\, U(\vek{R}) = \Pi^a_i(\vek{x}) + \epsilon^a_i(\vek{x}) \,,
\label{T1}
\end{equation}
where a new color electric flux field
\be
\Vek{\epsilon}^a(\vek{x}) \equiv \int\limits_{\mathscr{C}\,:\,
\mathbf{0} \to \mathbf{R}} d\vek{y}\,\delta(\vek{x}-\vek{y}) \, i
U(\vek{y})^\dagger\,T^a\,U(\vek{y}) =  \int\limits_{\mathscr{C}\,:\,
\mathbf{0} \to \mathbf{R}} d\vek{y}\,\delta(\vek{x}-\vek{y}) \, i
\widetilde{T}^a(\vek{y})
\label{epsilon}
\ee
with $\widetilde{T}^a(\vek{y}) \equiv U(\vek{y})^\dagger\,T^a\,U(\vek{y})$
has emerged.
The unitary transformation (\ref{T1}) removes the parallel
transporters from the Wilson state so that the Wilson loop becomes the expectation
value of the transformed time evolution operator in the Yang-Mills (zero charge)
vacuum,
\be
\langle\,W\,\rangle = \tr\,\langle \,0\,|\,e^{-T(
\widetilde{H}-E_0)}\,|\,0\,\rangle\,.
\label{W10}
\ee

The effect of the Wilson lines is now shuffled into the kinetic term of
the new Hamiltonian,
\begin{equation}
\widetilde{H} = \frac{g^2}{2} \int d^3x\, \Big[\Vek{\Pi}^a(\vek{x})\,\mathbbm{1}
+ \Vek{\epsilon}^a(\vek{x})\Big]^2 +
\frac{1}{2 g^2} \int d^3 x\, \mathbf{B}(\vek{x})^2\,\mathbbm{1}\,. \hk
\label{Htilde}
\end{equation}
Note that $\Vek{\epsilon}^a(\vek{x})$ depends on the gauge field, so that
$[\Pi^a(\mathbf{x}), \epsilon^b(\mathbf{x})] \neq 0$ which complicates
the treatment of the new  Hamiltonian considerably. The induced electric field
satisfies Gauss' law for the external charge density 
induced by the Wilson line,
\begin{equation}
[\, \hat{\vek{D}}^{ab}(\vek{x})\,,\,\Vek{\epsilon}^b(\vek{x})\,]
= U(\vek{R})^\dagger\,\rho^a(\vek{x}) \,U(\vek{R})
= i T^a\,\delta(\vek{x}) - i \widetilde{T}^a(\vek{R})\,\delta(\vek{x}-\vek{R})
\equiv \widetilde{\rho}^a(\vek{x})\,.
\label{rhotilde}
\end{equation}
This charge density $\widetilde{\rho}(\vek{x})$ is again a color matrix 
similar to eq.~(\ref{gauss2}).

\bigskip
Before proceeding, let us briefly discuss  the piece in the
transformed Hamiltonian (\ref{Htilde}) which is quadratic in the induced
electric field,
\begin{equation}
V_{\rm ind}[\mathscr{C}] = \frac{g^2}{2}\int d^3x\,\big[
\Vek{\epsilon}^a(\vek{x}) \big]^2\,.
\label{staticpot}
\end{equation}
After a simple calculation using the definition of the quadratic Casimir operator
$ \widetilde{T}^a\,\widetilde{T}^a = T^a\,T^a = C_2\,\mathbbm{1}$,
we find
\be
\label{4X-35}
V_{\rm ind} [\mathscr{C}]  = \frac{g^2}{2} \,C_2 \il_{\mathscr{C}\,:\,\vek{0} \to \vek{R}}
d \vx \, \il_{\mathscr{C}\,:\,\vek{0} \to \vek{R}} d \vy \,\delta^3 (\vx - \vy)
\,\mathbbm{1}\hk .
\ee
This term is independent of the gauge field and thus also of the vacuum wave
functional used, but it does depend on the shape of the contour
$\mathscr{C}\,:\,\mathbf{0} \to \mathbf{R}$ in the spatial part of the Wilson
loop. Choosing $\mathscr{C}$ to be a straight line connecting $\vek{0}$ and
$\vek{R}$, we have
\be
\label{F5X1}
V_{\rm ind} [\mathscr{C}] =  \frac{g^2}{2} \,C_2\,\delta^2_\perp (0)
\cdot | \vR| \,\mathbbm{1}\hk \,.
\ee
The color unit matrix disappears when the trace of the Wilson loop is taken.
While eq.~(\ref{F5X1}) is formally a linearly rising potential,
the corresponding string tension
\be
\label{5X-37}
\frac{g^2}{2} \,C_2 \,\delta^2_\perp (0) = \frac{\Lambda^2}{8 \pi} \,g^2 \,C_2
\ee
is UV-divergent (cutoff $\Lambda$) and of purely kinematic origin. Except for the Casimir
operator $C_2 \to 1$, the same contribution is also found in the case of QED which
does not confine; so it must be spurious. In fact, the singular energy is due to the
infinitely thin Wilson lines whose point-like cross section gives rise to the factor
$\delta^2_\perp (0)$ in eq.~(\ref{5X-37}). If the lines are smeared out to an effective
tube with cross section $\sim\Lambda^{-2}$, the self-energy is regularized,
$\delta^2_\perp (0) \sim \Lambda^2$, but it will still obscure the physical potential
unless it is properly isolated and removed.

To see the origin of the problem in more detail, consider the Abelian case
where Gauss' law for the induced electric field, $\nabla\Vek{\epsilon} = \rho$,
indicates that only the \emph{longitudinal} part of $\Vek{\epsilon}$ is
sensitive to external charges. As a consequence, the static potential
(\ref{staticpot}) decomposes into two parts,
\[
V_{\rm ind}[\mathscr{C}] = \frac{g^2}{2}\,\int d^3x\,\big[\Vek{\epsilon}_\perp(\vek{x})\big]^2
+ \frac{g^2}{2}\,\int d^3x\,\big[\Vek{\epsilon}_{\|}(\vek{x})\big]^2
\equiv V_\perp[\mathscr{C}] + V_{\|}[\mathscr{C}]\,.
\]
The first piece $V_\perp$ is built from $\Vek{\epsilon}_\perp$ alone and thus
insensitive to external charges; it represents a vacuum property that
renormalizes the Wilson loop but does not affect the physical potential between
static quarks. The latter is given exclusively by the second piece $V_{\|}$
which will be shown below to describe the well-known Coulomb interaction.
The separation of transversal and longitudinal degrees of freedom is thus
essential to disentangle the physical potential from renormalization effects.
In ref.~\cite{Gaete:1998vr}, a different approach to remove the self-energies
of the Wilson line and to isolate the proper quark potential
was proposed, which is based on partial wave expansion. Appendix \ref{appA} 
compares this method to our reasoning based on Gauss' law.

\subsection{Coulomb gauge}
\label{sec:coulomb}

Gauge invariant calculations of the Wilson loop are only possible in
special cases where the exact ground state is known analytically.
In realistic cases, the vacuum wave functional is unknown and one has to resort
to approximate calculations which usually require some sort of gauge fixing.
For the Hamiltonian approach, \emph{Coulomb gauge} is a natural choice which
simplifies the separation of dynamical from gauge degrees of freedom. 
In QED, for instance, the Coulomb gauge fixed field represents the 
physical (gauge invariant) degrees of freedom.

Not surprisingly, the variational approach to YM theory in Coulomb gauge has
recently become an excellent tool to obtain approximate ground state wave
functionals which are suitable for practical computations. We will use this
approach for YM-theory in $D=3+1$ in section \ref{sec:YM31}.

In view of our previous analysis, we can choose to fix the Coulomb gauge
either before or after the unitary transformation (\ref{T0}) that led from
expression (\ref{G8}) to eq.~(\ref{W10}). That is, starting from the invariant
Hamiltonian $H$ in eq.~(\ref{G2-X2}), we can either
\begin{enumerate}
\item \underline{gauge fix $H$ directly and compute $\langle W \rangle $
from eq.~(\ref{G8}): $\quad\langle W \rangle = \mathrm{tr}\,\left\langle \,
\mathbf{R}\,\left|\,e^{-T (H-E_0)}\,\right|\,\mathbf{R}\,\right\rangle$}

\smallskip\noindent
The gauge fixed Coulomb Hamiltonian can be obtained either by canonical or
by path integral techniques. The well-known result
is the expression \cite{Christ:1980ku}
\begin{eqnarray}
H &=& \frac{g^2}{2}\,\int d^3x\, \mathscr{J}^{-1}\,\Vek{\Pi}^\perp_a(\vek{x})
\,\mathscr{J}\,\Vek{\Pi}^\perp_a(\vek{x}) +
\frac{1}{2 g^2}\,\int d^3x\,\big[\,\vek{B}^a(\vek{x})\,\big]^2 + H_C
\label{Hfix}
\\[2mm]
H_C &=& \frac{g^2}{2}\,\int d^3(x,y)\,\mathscr{J}^{-1}\,\big[\rho^a(\vek{x}) +
\rho^a_{\rm dyn}(\vek{x})\big]\,
\mathscr{J}\cdot F^{ab}(\vek{x},\vek{y})\,\big[\rho^b(\vek{y}) +
\rho_{\rm dyn}^b(\vek{y})\big]\,,
\label{HC}
\end{eqnarray}
where $\mathscr{J}[A] = \det(-\hat{\vek{D}}\nabla)$ is the
Faddeev--Popov determinant and the Coulomb kernel $F$ reads
\[
F = (-\hat{\vek{D}}\nabla)^{-1}(-\Delta)\,(-\hat{\vek{D}}
\nabla)^{-1}\,.
\]
The resolution of Gauss' law must take into account that the state
$|\,\vek{R}\,\rangle$ in eq.~(\ref{G8}) is charged, i.e.~besides the
dynamical charge of the gluons,
$\rho_{\rm dyn}^a(\vek{x}) \equiv - \hat{\vek{A}}^{ab}_\perp(\vek{x})\,
\Vek{\Pi}^b_\perp(\vek{x})$ (with $\hat{A}_\perp^{ab} \equiv f^{acb}\,A_\perp^c$),
we also have to include the external charge $\rho^a$ induced by the Wilson
line. For instance, from Gauss' law in the form eq.~(\ref{gauss2})
\[
\big( \hat{\vek{D}}^{ab}\,\Vek{\Pi}^b\big)\,\Psi_{ij} = - \rho^a_{ik}\,\Psi_{kj}
\qquad \Longrightarrow\qquad \nabla \Vek{\Pi}^a\,\Psi_{ij} = -
\big[ \rho^a + \rho_{\rm dyn}^a\,\mathbbm{1}\big]_{ik}\,\Psi_{kj}
\]
the usual resolution entails that the charges in eq.~(\ref{HC}) are really
matrices $[\rho^a + \rho^a_{\rm dyn}\,\mathbbm{1}]$, and so is the
Coulomb kernel $H_C$.  The Wilson state $|\,\vek{R}\,\rangle$,
after gauge fixing, depends on $\vek{A}^\perp$ only and is thus no longer from the
$q\bar{q}$-sector; it will, in fact, have overlap with the zero-charge sector.

\item \underline{compute $\widetilde{H} = U(\mathbf{R})\,H\,U(\mathbf{R})^\dagger$ and
gauge fix expression eq.~(\ref{W10}):\quad $\langle W \rangle = \langle
\,0\,|\,\tr\,e^{-T (\widetilde{H}-E_0)}\,|\,0\,\rangle $}

\smallskip\noindent
Since the Wilson lines have been taken care of by the unitary transformation
$H \to \widetilde{H}$ in eq.~(\ref{T0}), no further external charges have to be 
included, and Gauss' law takes the form 
$\hat{\vek{D}}\hat{\Vek{\Pi}}\,|\,0\,\rangle = 0$.
All calculations can thus be done in the zero-charge sector.
From eq.~(\ref{Htilde}), we could now follow the standard route and apply the
Faddeev--Popov procedure to the vev
\[
\langle\,0\,|\,\widetilde{H}\,|\,0\,\rangle = \int \mathscr{D}\vek{A}\,
\big[ (\Vek{\Pi} + \Vek{\epsilon})\,\Psi_0 \big]^\ast\,\big[(\vek{\Pi} +
\vek{\epsilon}) \,\Psi_0 \big] + \cdots\,
\qquad\quad \Psi_0[\vek{A}] \equiv \langle\, \vek{A}\,|\,0\,\rangle \,.
\]
Using Gauss' law to solve for $\vek{\Pi}^\|\,\Psi_0$ eliminates all
longitudinal d.o.f.~and yields the vev
$\langle\,0\,|\,\widetilde{H}^\perp\,|\,0\,\rangle$ in terms of transversal
fields only. However, this is usually not sufficient to find the required
vev of the gauge fixed \emph{evolution operator}
$\langle\,0\,|e^{- T \widetilde{H}^\perp}\,|\,0\,\rangle$, except for
special cases such as QED.
\end{enumerate}

Thus for the non-Abelian case, we prefer to follow the first method based on
the standard Coulomb gauge Hamiltonian, while Abelian systems can be treated
with both approaches.

\section{Quantum electrodynamics}
\subsection{Gauge invariant treatment}
\no
In the Abelian case $G=U(1)$, the Maxwell gauge potentials do not carry a
color index, and the hermitian generator of the non-Abelian
gauge group, $(iT^a)$, has to be replaced by $1$, so that also
$C_2 = 1$. The formula for the induced electric field in the representation
eq.~(\ref{Htilde}) simplifies to
\begin{equation}
\Vek{\epsilon}(\vek{x}) = \int\limits_{\mathscr{C}}
d\vek{y}\,\delta^{(3)}(\vek{x}-\vek{y})\,.
\end{equation}
In particular, $\Vek{\epsilon}(\vek{x})$ is field-independent and
the parallel transporter $\vek{0} \to \vek{R}$ becomes
\[
U(\vek{R}) = \exp\left[ i \int d^3 x\,\Vek{\epsilon}(\vek{x})\cdot
\vek{A}(\vek{x})\right].
\]
To isolate the physical degrees of freedom, we split all vector fields in
longitudinal and transversal parts. Without gauge fixing, the ground state
$\Psi_0[\vek{A}] = \langle\,\vek{A}\,|\,0\,\rangle$ is
eigenstate of the gauge invariant Hamiltonian, and thus gauge invariant itself; as
a consequence, it can only depend on the physical (i.e.~transversal)
field~$\vek{A}^\perp$\,, so that $\Vek{\Pi}^\|\,|\,0 \,\rangle = 0$.
From eq.~(\ref{W10}) and (\ref{Htilde}), this entails
\begin{eqnarray}
\langle\, W \,\rangle &=& \left\langle \, 0 \,\left|\,e^{-T\left(
\widetilde{H}_{\rm QED}^\perp - E_0\right)}\,\right|\,0\,\right\rangle
\nonumber \\[4mm]
\widetilde{H}_{\rm QED}^\perp &=&
\frac{1}{2} \int d^3 x \left[ g^2\,\big[ \vek{\Pi}^\perp
(x) + \Vek{\epsilon}^\perp (x) \big]^2 + \frac{1}{g^2}\,
\vB^2 (x) + g^2\,\big[\Vek{\epsilon}^{\|}(x)\big]^2
\right]\,. \hk
\label{HQEDtildeperp}
\end{eqnarray}
Since $\Vek{\epsilon}$ is independent of the gauge field, the last term
in $\widetilde{H}_{\rm QED}^\perp$ commutes with the remainder and can be
pulled out of the
\emph{vev},
\begin{eqnarray}
\langle \, W\,\rangle &=& e^{- T V^{\|}}\cdot \left\langle \,0\,\left|\,
e^{- T \left( \widetilde{H}_{\rm QED}' -E_0\right)}
\,\right|\,0\,\right\rangle \nonumber \\[2mm]
V^{\|} &=& \frac{g^2}{2}\,\int d^3 \vek{x}\,\big[ \Vek{\epsilon}^\|(\vek{x})\big]^2\,,
\end{eqnarray}
where $V_{\|}$ is the part of the induced static potential that is
sensitive to external charges, cf.~our previous discussion.
The remaining Hamiltonian
\[
\widetilde{H}_{\rm QED}' = \frac{1}{2} \int d^3 x \left[ g^2\,\big[
\vek{\Pi}^\perp (\vek{x}) + \Vek{\epsilon}^\perp (\vek{x}) \big]^2
+ \frac{1}{g^2}\,\big[\nabla \times \vek{A}^\perp(\vek{x})\big]^2
\right]
\]
depends on the physical degrees of freedom $\vek{A}^\perp$ only. It is most
easily treated by reversing the transformation eq.~(\ref{T1}), but this time
with the parallel transporter made of physical gauge fields $\vek{A}^\perp$
only,
\begin{equation}
U^\perp(\vek{R}) \equiv \exp\left[ i \int_{\mathscr{C}\,:\,\vek{0} \to \vek{R}}
d\vek{x}\,\vek{A}^\perp(\vek{x})\right] =
\exp\left[ i \int d^3 \vek{x}\,\Vek{\epsilon}^\perp(\vek{x}) \cdot
\vek{A}^\perp(\vek{x})\right]\,.
\label{KH1}
\end{equation}
As a result, the induced transversal electric field $\Vek{\epsilon}^\perp$
is removed from the Hamiltonian and re-shuffled into the state
$|\,\vek{R}^\perp\,\rangle = U^\perp(\vek{R})\,|\,0\,\rangle$. Although this
new state $|\,\vek{R}^\perp\,\rangle$ resembles the Wilson state, it depends on
$\vek{A}^\perp$ only and is thus gauge invariant, i.e.~$|\,\vek{R}^\perp\,\rangle$
lies in the zero-charge sector.
Inserting a complete set of zero-charge (gauge invariant) eigenstates of the standard
QED Hamiltonian,
\begin{equation}
H_{\rm QED} = \frac{1}{2} \int d^3 x \left[ g^2\,\big[
\vek{\Pi}^\perp (\vek{x}) \big]^2 + \frac{1}{g^2}\,\big[\nabla
\times \vek{A}^\perp(\vek{x})\big]^2 \right]
\label{KH2}
\end{equation}
with $\Vek{\Pi}^\perp = \delta / i\delta \vek{A}^\perp$ we find
 \be
 \label{G8-74}
 \langle \,W\, \rangle = e^{- T V_{\|}} \,\sli_n e^{- \lk E_n - E_0 \rk T} \,
 \left|\, \left\langle \,n \,\left |\, U^\perp(\vek{R})
 \,\right|\, 0 \,\right \rangle\, \right|^2 \hk \,.
 \ee
It is clear that only the first term ($n=0)$ in the sum survives in the limit
$T \to \infty$, provided that the matrix element
$\langle \,0\,|\,U^\perp(\mathbf{R})\,|\,0\,\rangle = \langle\,0\,|\,\vek{R}^\perp\,\rangle$
is non-vanishing. In the present case, this matrix element can be computed
explicitly, since the vacuum wave functional is Gaussian,
\[
\Big\langle\,\vek{A}^\perp(\vek{x})\,\Big|\,0\,\Big\rangle =
\mathscr{N}\,\exp\left[ - \frac{1}{g^2}\,\int d^3(x,y)\,
\vek{A}^\perp(\vek{x})\,\omega(\vek{x},\vek{y})\,\vek{A}^\perp(\vek{y})\right]\,.
\]
A simple Gaussian integral gives
\begin{eqnarray}
\langle\,0\,|\,U^\perp(\vek{R})\,|\,0\,\rangle &=&
\left\langle\,0\,\left|\,\exp\left[i \int d^3x\,\vek{A}^\perp(\vek{x})\cdot
\Vek{\epsilon}^\perp(\vek{x})\right] \,\right|\,0\,\right\rangle
= \nonumber \\[2mm]
&& \hspace*{-2cm} =\,\,\,\exp\left[-\frac{g^2}{4}\,\int d^3(x,y)\,\Vek{\epsilon}^\perp(\vek{x})\,
\omega^{-1}(\vek{x},\vek{y})\,\Vek{\epsilon}^\perp(\vek{y})\right] \nonumber
= \exp\left[-\frac{g^2}{4}\,\int_\vek{0}^\vek{R} dx_i\int_\vek{0}^\vek{R}
dy_j\,t_{ij}(\vek{x})\,\omega^{-1}(\vek{x},\vek{y})\right]\,,
\nonumber
\end{eqnarray}
where $t_{ij}$ is the usual transversal projector. This expression
can be further evaluated if we choose the path $\vek{0} \to \vek{R}$ to be a
straight line and employ the QED dispersion relation $\omega(\vek{k}) = |\vek{k}|$
to invert the kernel $\omega(\vek{x},\vek{y})$ in momentum space.
Introducing spherical coordinates in momentum space, the parameter integrals for
the Wilson lines can be done first, while the subsequent angle integral can be
expressed through the integral sinus, $\mathrm{Si}(x)$. For the remaining
integral over $k = |\vek{k}|$, we use a sharp UV cutoff (i.e.~a $O(3)$ invariant
cutoff in 3-space), to find
\begin{eqnarray}
\langle\,0\,|\,\vek{R}^\perp\,\rangle &=&
\exp\left[ - \frac{g^2}{8 \pi^2}\left( 4 \gamma + 2 \cos (\Lambda R) - 4
\ln (\Lambda R) +
 4 \,\mathrm{Ci} (\Lambda R) - 2\, \frac{\sin (\Lambda R)}{\Lambda R} + 2 \,\Lambda R\,
 \mathrm{Si} (\Lambda R)\right)\,\right] \nonumber\\[2mm]
& = & \exp\left[- \frac{g^2}{8 \pi^2} \left( \pi \Lambda R -
4 \ln (\Lambda R) - 4 \gamma + \mathscr{O}(\Lambda^{- 3})\right)\,\right]\,.
\label{QED_overlap}
\end{eqnarray}
In the limit $T \to \infty$, the bare Wilson loop thus becomes
\begin{equation}
\langle \,W\,\rangle = Z_W(\Lambda R)\cdot \exp\left[- T\,V_{\|}(R)\right]
\label{Wbare}
\end{equation}
where
\[
Z_W(x) =\exp\left[ - \frac{g^2}{4 \pi^2}\,\left(\pi x - 4 \ln x -4
\gamma\right)\right]\,.
\]
Formally, $Z_W$ vanishes in the limit $\Lambda \to \infty$, which indicates that
the state $|\,\vek{R}^\perp\,\rangle$ with support on an infinitely thin line
has poor overlap with the true QED ground state. On the other hand,
$Z_W$ is independent of $T$ and therefore does not contribute to the potential.
The correct interpretation is given by the operator product expansion (OPE)
\cite{Collins:1984xc}:
Since the Wilson loop $W$ is a composite operator involving products
of arbitrarily many gauge fields, we expect short-distance OPE divergences associated with
the operator $W$, in addition to those counter terms which are necessary to render
elementary Green functions finite. Eq.~(\ref{Wbare}) shows that this is
indeed the case, with one overall multiplication of the operator
$W \to W_R \equiv Z_W^{-1}\cdot W$ rendering all matrix elements finite.
This renormalization property of Wilson loops is an exact result that has been
known for a long time both in the Abelian and non-Abelian case
\cite{Gervais:1979fv, Polyakov:1980ca}. From the renormalized value
\[
\langle\,W_R\,\rangle = \exp\left[- T \,V^{\|}(R)\right]\,,\qquad\quad T \to \infty
\]
we conclude that, in the Abelian case, the Wilson potential agrees with 
$V_{\|}$. A simple explicit calculation gives first the induced electric field
\begin{eqnarray}
\Vek{\epsilon}^\|(\vek{x}) &=& - \nabla_x\,(-\Delta_x^{-1})\,\nabla_x \,
\Vek{\epsilon}(\vek{x}) = \nabla_x\,(-\Delta_x^{-1})\,\int_\vek{0}^\vek{R}
dy\,\nabla_{y}\,\delta(\vek{x}-\vek{y})
\nonumber \\[2mm]
&&\qquad\qquad = \,\,\,- \nabla_x \,(-\Delta_x^{-1})\,
\left[ - \delta(\vek{x}-\vek{R}) + \delta(\vek{x}) \right]
\equiv - \nabla_x \,(-\Delta_x^{-1})\,\rho(\vek{x})
\end{eqnarray}
and thus Gauss' law $\nabla\Vek{\epsilon}^\| = \rho$. The Wilson
potential becomes, after integration by parts and using the Green function
$G(\vek{x},\vek{y}) = \langle\,\vek{x}\,|\,(-\Delta)^{-1}\,|\,\vek{y}\,\rangle
=(4 \pi |\vek{x}-\vek{y}|)^{-1}$
\begin{eqnarray}
V^\|(R) &=& \frac{g^2}{2}\,\int d^3\vek{x}\,\Big[
\Vek{\epsilon}^\|(\vek{x})\Big]^2
= \frac{g^2}{2}\int d^3(\vek{x},\vek{y},\vek{z})\Big[ - \nabla_x\,G(\vek{x},\vek{y})\,\rho(\vek{y})
\Big]\cdot\Big[ - \nabla_x\,G(\vek{x},\vek{z})\,\rho(\vek{z}) \Big]
= \nonumber \\[2mm]
&=& \frac{g^2}{2}\int d^3(\vek{x},\vek{y})\,\rho(\vek{x})\,G(\vek{x},\vek{y})\,
\rho(\vek{y}) = \frac{g^2}{2}\,\int d^3(\vek{x},\vek{y}) \,
\frac{\rho(\vek{x})\,\rho(\vek{y})}{4\pi \,|\vek{x}-\vek{y}|} =
\nonumber \\[2mm]
&=&\frac{g^2}{4 \pi |\vek{R}|} - \frac{g^2}{ 4 \pi \,|\vek{0}|}\,.
\label{WQED}
\end{eqnarray}
This is just the familiar Coulomb law including  the self-energy of the
two charges.

\subsection{Coulomb gauge}
In the Abelian case, we can apply both methods described in section
\ref{sec:coulomb} to evaluate the Wilson loop in Coulomb gauge:
\begin{enumerate}
\item In the Abelian case, the gauge fixed Hamiltonian eq.~(\ref{Hfix}) simplifies
considerably. From $\mathscr{J} = \mbox{const}$, $F = (-\Delta)^{-1}$ and
$\rho_{\rm dyn} = 0$, we have
\[
H = \frac{1}{2} \int d^3 x \left[ g^2\,
\vek{\Pi}^\perp (\vek{x})^2+ \frac{1}{g^2}\,\vek{B}(\vek{x})^2 \right] +
\frac{g^2}{2} \int d^3 (x,y)\,\rho(\vek{x}) (-\Delta)^{-1}(\vek{x},\vek{y})\,
\rho(\vek{y}) \equiv H_{\rm QED} + H_C\,.
\]
The second piece $H_C$ is a field-independent $c$-number, so that
$e^{- T H} = e^{- T\,H_{\rm QED}} \cdot e^{- T\,H_C}$. Taking into account
that $|\,\vek{R}\,\rangle = U[\vek{A}^\perp]\,|\,0\,\rangle$ and
inserting a complete set of eigenstates of $H_{\rm QED}$, we arrive directly
at eq.~(\ref{G8-74}), from which we can follow all the steps to the final
expression eq.~(\ref{WQED}) for the Wilson loop.
\item Resolution of Gauss' law in the chargeless sector gives
$\Vek{\Pi}^\|\,|\,0\,\rangle = 0$, and after splitting into longitudinal and
transversal fields and Coulomb gauge fixing, the operator $\widetilde{H}$
in eq.~(\ref{Htilde}) becomes $\widetilde{H}^\perp_{\rm QED}$ in
eq.~(\ref{HQEDtildeperp}). From there, we can again follow all the steps that
led eventually to the expression eq.~(\ref{WQED}) for the Wilson loop.
\end{enumerate}

\section{Yang-Mills theory in $D=1+1$}
\label{sec:YM11}
\no
In the non-Abelian case, the exact vacuum wave functional is only known for
the special case of one space dimension, which is nonetheless a good testing
ground since the exact result for the string tension is known e.g.~from path
integral techniques \cite{Witten:1991we,Blau:1991mp}.
However, YM theory in $D=1+1$ is (almost) topological and non-trivial results can
only be obtained on non-contractible Euclidean spacetime-manifolds. Our Hamiltonian
approach at zero temperature requires that Euclidean time be unrestricted, so that
the only non-trivial spacetime manifold is $\mathsf{M} = \mathbbm{R} \times
S^1$, where the first factor is time and the second factor is space. In the
following, we will therefore assume that the spacetime is a cylinder, i.e.~space
is the interval $[0,L]$ with periodic boundary conditions.
(The general theory of fiber bundles indicates that transition functions on
$\mathsf{M}$ can be taken trivial, i.e.~gauge fields are strictly
periodic while gauge transformations are periodic up to a center element.)

\subsection{Gauge invariant treatment}
In $D=1+1$, there is no magnetic field and the vacuum is characterized by
the condition $\Pi^a(x)\,\big|\,0\,\big\rangle = 0$. This is because then
$\langle\,0\,|\,H\,|\,0\,\rangle = 0$ which in view of $H \ge 0$ is an
absolute minimum so that $|\,0\,\rangle$ must be the ground state with zero
energy $E_0 = 0$. The condition $\Pi^a\,\big|\,0\,\big\rangle = 0$ can be
readily solved by a constant 
\begin{equation}
\Psi_0[A] = \langle\,A\,|\,0\,\rangle = \mbox{const}\,.
\end{equation}
If we first ignore the commutator $[\Pi^a(\vek{x}),\epsilon^b(\vek{x})] \neq 0$
in eq.~(\ref{Htilde}), we find immediately $\widetilde{H}\,|\,0\,\rangle =
V_{\rm ind}[\mathscr{C}]\cdot|\,0\,\rangle$, where $V_{\rm ind}[\mathscr{C}]$
is the static potential defined in eq.~(\ref{4X-35}). As a consequence, we have
\[
\langle \, W\,\rangle = \tr\,
\big\langle\,0\,\big|\,e^{- T \widetilde{H}}\,
\big|\,0\,\big\rangle = \mathrm{tr}\,e^{- T V_{\rm ind}[\mathscr{C}]}
\]
i.e.~the Wilson potential matches the induced potential eq.~(\ref{4X-35}),
\begin{equation}
V(R) = V_{\rm ind}[0 \to R] = \frac{g^2}{2}\,C_2\,\mathbbm{1}\,
\int_0^R dx\int_0^R dy\, \delta(x,y) = \frac{g^2}{2}\,C_2\, R\,\mathbbm{1}.
\label{pot11}
\end{equation}
We are thus led to a linearly rising potential with the string tension
\begin{equation}
\sigma_{1+1} = \frac{g^2}{2}\,C_2\,.
\label{sigma11}
\end{equation}
This is the correct result known e.g.~from path integral techniques 
\cite{Witten:1991we,Blau:1991mp} or from the lattice Hamiltonian 
\cite{Burgio:1999tg}. However, there are two caveats to this result:
\begin{itemize}
\item we have neglected the commutator $[\Pi^a(x),\epsilon^a(x)]$ in the
Hamiltonian ;
\item we have not yet made use of the periodic boundary conditions and the
resulting potential eq.~(\ref{pot11}) is \emph{not} invariant under the
replacement $R \to L-R$, nor is it periodic in $R$.
\end{itemize}
We will not try to solve the first problem here, since the correct treatment of
the dynamics will be easier in Coulomb gauge as described in the next section.
As for the periodicity, the distances $R$ and $L-R$ are indeed equivalent on a
spacetime cylinder, but the Wilson loop will \emph{not} be invariant under the
replacement $R \to L-R$. This is not a contradiction since the physical potential
in the compact spacetime setting is no longer given by the exponential of any
single Wilson loop.

\begin{figure}
\includegraphics[width = 8cm]{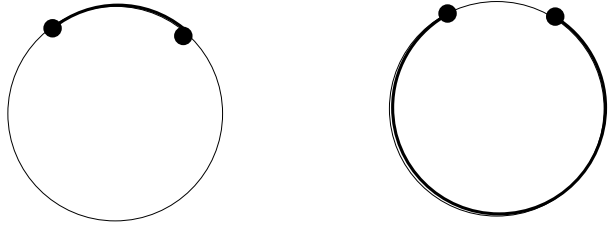}
\caption{Two possible Wilson lines connecting the same points on the compact
space interval with periodic boundary conditions.}
\label{fig:wilson11}
\end{figure}

To understand this subtle point in more detail, we have to go back to the
physical interpretation of the Wilson loop in section \ref{general}.
The identification of the (exponent of the) Wilson loop with the static
quark potential relies on the assumption that the Wilson state made by
the direct path $\mathscr{C}\,:\,0 \to R$ has non-vanishing overlap with the
true YM ground state in the $q\bar{q}$-sector for all values of $R$. However,
the compact space interval allows for many more Wilson states from the same
$q\bar{q}$-subspace, namely (i) the state where the path $\mathscr{C}$ winds
around the compact space $S^1$ in the opposite direction (cf.~figure \ref{fig:wilson11}),
and (ii) all states which can be obtained by winding the Wilson line an additional
integer number of times around the space in any direction. We can label all
these states by the end point of the Wilson line used to create them,
$\big\{ |\,R + m L \,\rangle\,;\,m \in \mathbbm{Z}\big\}$. (A
negative end point indicates that the Wilson line winds around space in negative
direction; for instance, the two lines in figure \ref{fig:wilson11} correspond
to the states $|\,R\,\rangle$ and $|\,R-L\,\rangle$, respectively.)

At small $R \ll L$, we expect that the state $|\,R\,\rangle$ is energetically
favored, but this may change at larger $R$ so that the Wilson loop with the
direct path connecting $0 \to R$ will no longer give the correct quark potential.
Assuming that at least one of the Wilson states $|\,R + m L\,\rangle$ has
overlap with the true $q\bar{q}$ ground state, we can extract the true
quark potential from the largest eigenvalue of the matrix
\begin{equation}
\langle \, R + m L \, | \,e^{- T (H-E_0)}\,| R + n L \,\rangle\,,
\qquad\quad m,n \in \mathbbm{Z}\,.
\label{matrix11}
\end{equation}
The same information may also be obtained from the diagonal matrix elements,
\begin{equation}
V(R) = \min_{m \in \mathbbm{Z}} \Big[- \lim_{T \to \infty}T^{-1}\,\ln\,
\langle\,R + m L \,|\,e^{- T(H-E_0)}\,|\,R + m L\,\rangle\Big]\,.
\label{truepot11}
\end{equation}
This follows from section \ref{general}, because the quantity in square brackets
is simply $(E_{\ell(m)}^{(q\bar{q})} - E_0)$, where $E_{\ell(m)}^{(q\bar{q})}$ refers to the
lowest Yang-Mills eigenstate in the $q\bar{q}$-sector which has non-vanishing
overlap with the Wilson state $|\,R + m L\,\rangle$. Assuming that at least
one of these Wilson states has overlap with the true $q\bar{q}$-\emph{vacuum},
we must have $\ell(m)=0$ for at least one $m$. Thus taking the minimum with 
respect to $m$ yields $(E_0^{(q\bar{q})}-E_0)$ on the rhs of 
eq.~(\ref{truepot11}), which is the proper static quark potential. 
However, since the minimum can (and will be) be at different
$m$ for different distances $R$, the true quark potential is not obtained from
any single Wilson loop. It is already evident that the potential (\ref{truepot11})
has all the required periodicity properties; the explicit evaluation of the
matrix elements is, however, best performed in Coulomb gauge.

\subsection{Coulomb gauge}
Our treatment of Yang-Mills theory in $D=1+1$ in Coulomb gauge will follow
the first method outlined in section \ref{sec:coulomb}, i.e.~we gauge fix the
usual Yang-Mills Hamiltonian directly, before removing the Wilson line from the
states.
In $(1 + 1)$ dimensions, the  Coulomb condition
\be
\label{16-146}
\partial_1 A^a_1 = 0
\ee
leaves a (spatially) constant gauge field $A^a_1 = \mathrm{const}$, which can be
further diagonalized by means of the residual global color symmetry 
(\emph{diagonal Coulomb gauge}) \cite{Reinhardt:2008ij, Hetrick:1993tv}. 
For simplicity, we will restrict the color group in
this subsection to $G=SU(2)$ with the antihermitean generators
$T^a = \tau^a / (2i)$, where $\tau^a$ are the Pauli matrices. In the
diagonal Coulomb gauge, we are then left with only one physical degree of
freedom, $A_1^{a=3}$, which we normalize to the dimensionless variable
\be
\vartheta = \frac{1}{2}\,A_1^3\,L \in [0,\pi]\,.
\label{theta}
\ee
The restriction to the compact range --- the fundamental modular region ---
ensures uniqueness of the gf.~section and avoids the Gribov problem. Using this
variable, the gauge-fixed Hamiltonian in the absence of external charges
becomes \cite{Reinhardt:2008ij}
\be
\label{G50}
H = - \frac{1}{2 L} \lk \frac{g L}{2} \rk^2 \frac{1}{\sin^2 \vartheta}
\frac{d}{d \vartheta} \sin^2 \vartheta \frac{d}{d \vartheta} \hk .
\ee
The corresponding eigenfunctions $H\,|\,n\,\rangle = E_n\,|\,n\,\rangle$
are just the characters of $SU(2)$ and thus known explicitly,
\begin{equation}
\langle\,\vartheta\,|\,n\,\rangle = \sqrt{\frac{2}{\pi}}\cdot\frac{\sin
(n+1)\vartheta}{\sin\vartheta}\,,\qquad\quad
E_n = \frac{1}{2L}\,\left(\frac{gL}{2}\right)^2\,n(n+2)\,,\qquad\quad
n \in \mathbbm{N}\,.
\label{G51}
\end{equation}
In \emph{diagonal} Coulomb gauge, both $\vek{A}$ and $\Vek{\Pi}$ live in the
Cartan algebra, so that their color commutator (the gluon charge) $\rho_{\rm dyn}$
vanishes. As a consequence, the Coulomb term $H_C$ from eq.~(\ref{HC})
vanishes completely  unless \emph{external}
charges $\rho \neq 0$ are present. For the Wilson loop, the external charge
$\rho$ is given by  eq.~(\ref{gauss2}); it is field dependent but contains no
derivative operator $\vek{\Pi}$ and thus commutes with the Jacobian
$\mathscr{J}$. We are thus left with
\be
H_C = \frac{g^2}{2} \int d(x,y)\,\rho^a (x) \,F^{a b} (x, y)\, \rho^b (x) \hk \,.
\label{HC11}
\ee
Notice that this expression differs from the static Coulomb potential because
the charges $\rho(x)$ from eq.~(\ref{gauss2}) still contain the parallel
transporters from the original Wilson lines.

\begin{figure}
\includegraphics[width = 7cm]{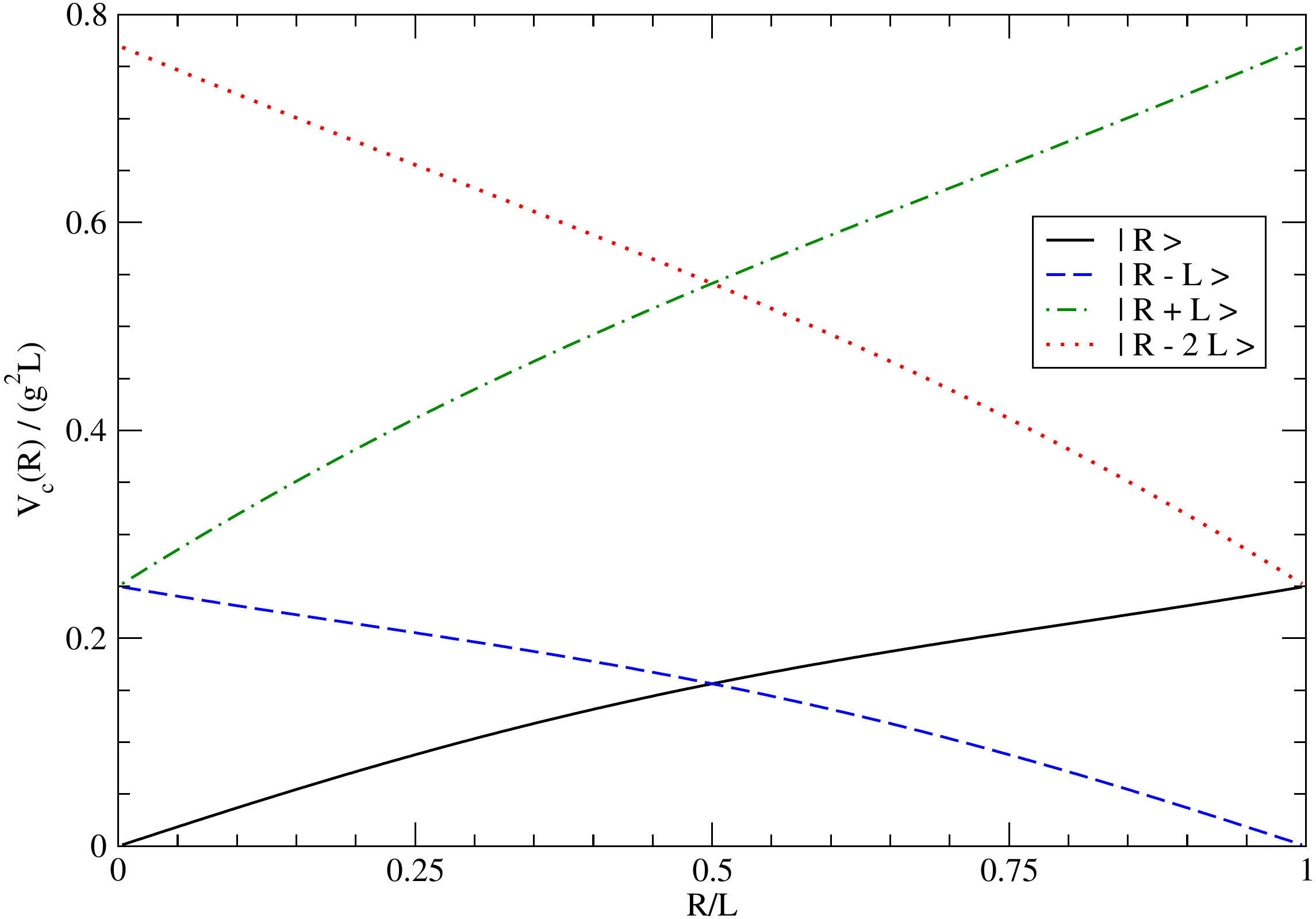} \hspace*{\fill}
\includegraphics[width = 7cm]{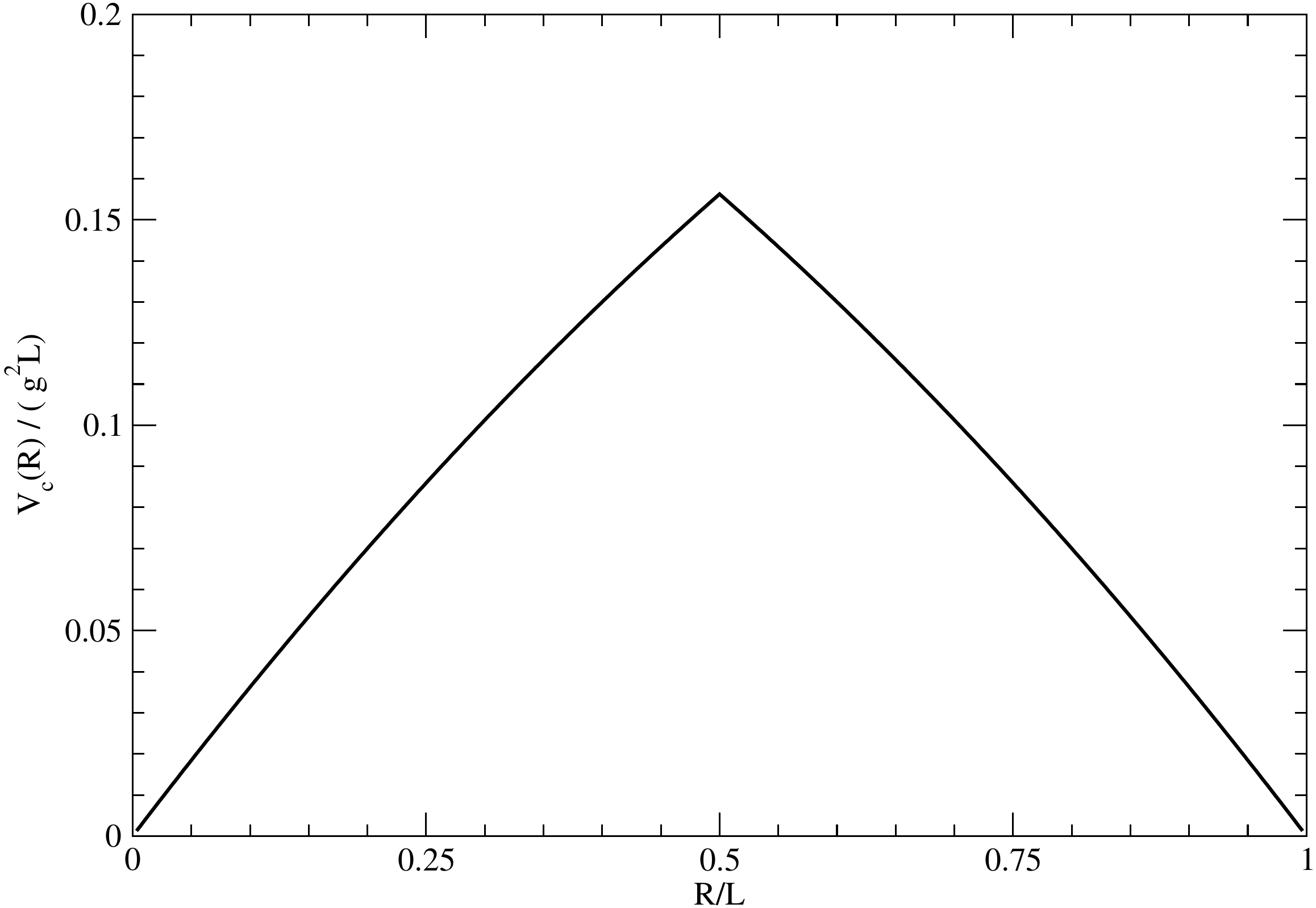}
\caption{Left panel: The Coulomb Hamiltonian $V_C(z)$ for the four shortest
Wilson states $|\,z\,\rangle$. Right panel: The true quark-antiquark potential for
$D=1+1$ Yang-Mills theory on the spacetime cylinder, as extracted from
the prescription eq.~(\ref{truepot11}).}
\label{fig:ym11}
\end{figure}

To further compute eq.~(\ref{HC11}), it is convenient to introduce a polar basis
in color space,
\be
\label{11-17}
\ve_{\sigma = 1} = - \frac{1}{\sqrt{2}} \lk \begin{array}{c} 1 \\ i \\ 0
\end{array} \rk \hk , \hk \ve_{\sigma = - 1} = \frac{1}{\sqrt{2}} \lk
\begin{array}{c} 1 \\ - i \\ 0
\end{array} \rk \hk , \hk \ve_{\sigma = 0} = \lk
\begin{array}{c} 0 \\ 0 \\ 1
\end{array} \rk \hk
\ee
and expand all relevant color vectors in this basis,
\begin{equation}
\begin{array}{r@{\,\,\,=\,\,\,}l@{\qquad}r@{\,\,\,}c@{\,\,\,}l}
\rho_\sigma & e_\sigma^a\,\rho_a & \rho_\sigma^\dagger &=& (e_\sigma^a)^\ast\,\rho_a \\[2mm]
\tau_\sigma & e^a_\sigma\,\tau_a & \tau_\sigma^\dagger &=& (e_\sigma^a)^\ast\,\tau_a
= (-1)^\sigma\,\tau_{-\sigma} \\[2mm]
(e_\sigma^a)^\ast\,F^{ab}(x,y)\,e_\tau^b & \delta_{\sigma\tau}\,F_\sigma(x-y)\,. &&&
\end{array}
\end{equation}
After a simple calculation, we find the rotated generators
\begin{equation}
\begin{array}{r@{\,\,\,=\,\,\,}l@{\qquad}r@{\,\,\,=\,\,\,}l}
U^\dagger \,\tau_\sigma \, U &  e^{-2i \alpha \sigma} \,\tau_\sigma
& U^\dagger\,\tau_\sigma^\dagger\,U &
e^{2i \alpha \sigma} \,\tau^\dagger_\sigma
\nonumber\\
U \,\tau_\sigma\, U^\dagger &  e^{2i \alpha \sigma} \,\tau_\sigma &
U \,\tau^\dagger_\sigma \,U^\dagger & e^{- 2i \alpha \sigma} \,\tau_\sigma \hk .
\end{array}
\label{11-21}
\end{equation}
where $U$ is the parallel transporter $0 \to R + m L$ in diagonal Coulomb gauge,
\begin{equation}
U = \exp\left(i \alpha\,\tau_3\right)\,,\qquad\quad \alpha =
\left(\frac{R}{L} + m\right)\,\vartheta\,.
\end{equation}
We can now insert this representation in eq.~(\ref{gauss2}) to find the external
charge of the Wilson state $|\,R + m L \,\rangle$ in the form
\begin{equation}
\rho_\sigma (x) = - \delta (x - R - m L) + \delta (x)\,
e^{2 i\sigma\left(\frac{R}{L} + m\right)\vartheta}\,.
\label{1608}
\end{equation}
As mentioned earlier, this differs from the sum of two static point charges by
the Wilson line connecting them. The Coulomb Hamiltonian eq.~(\ref{HC11})
can now be put in the form
\begin{eqnarray}
H_C &=& \frac{g^2}{8} \sum_{\sigma=-1}^{+1} h_\sigma (z) \,\tau_\sigma
\tau^\dagger_\sigma \equiv H_C(z)\nonumber \\[2mm]
h_\sigma (z) &\equiv& 2 F_\sigma (0) - \left[ e^{- 2i \sigma \frac{z}{L} \vartheta}
\,F_\sigma (z) + e^{2i \sigma \frac{z}{L} \vartheta} \,F_\sigma (-z) \right] \hk ,
 \label{1606}
\end{eqnarray}
where $z = R + m L$.
To complete the calculation of $H_C$, we can use the explicit expressions for
the Coulomb kernel derived in ref.~\cite{Reinhardt:2008ij} to find the 
polar basis representation
\bea
F_{0} (z) & = & \frac{L}{2 \pi^2}\sum_{n=1}^\infty
\frac{\cos(2 \pi n z/L)}{n^2} =
\frac{z^2}{2 L} - \frac{|z|}{2} + \frac{L}{12} \hk \nonumber\\[2mm]
\label{11-25}
F_{\pm} (R) & = & \frac{L}{4}\sum_{n=-\infty}^\infty \frac{e^{-2 \pi i n z/L}}
{(\pi n \pm \vartheta)^2} =
e^{2 i \sigma \frac{z}{L}\vartheta} \left[
\frac{L}{4 \sin^2 \vartheta} - \frac{|z|}{2} \mp i  \,\frac{z}{2} \,\cot
\vartheta \right] \hk\,.
\eea
The Fourier representation shows that the Coulomb kernel is indeed periodic,
$F_\sigma(R + m L) = F_\sigma(R)$, while the explicit expressions on the
rhs.~are only valid for $|z| \le R$, with periodic continuation outside this
range. After a brief calculation, we find, again in the range $|z| \le L$,
\be
\label{11-26}
h_0 (z) = | z | - \frac{z^2}{L}\,,\qquad\quad
h_{\pm} (z) = | z | \hk .
\ee
With $\tau_\sigma\tau_\sigma^\dagger = \mathbbm{1}  + \sigma\,\tau_3$,
the Coulomb term is finally given by
\be
\label{11-28}
H_C(z) = \frac{g^2}{8} \,\left[ 3 | z | - \frac{z^2}{L} \right]\,\mathbbm{1}\,,
\qquad\quad |z| \le L\,.
\ee
Notice that this formula is only valid for $|z| = | R + mL| < L$, i.e.~for the
two states $|\,R\,\rangle$ and $|\,R-L\,\rangle$ from figure \ref{fig:wilson11}.
For other Wilson states, we have to go back to the Fourier representation
in eq.~(\ref{11-25}) and do the sums numerically. It is then easy to see that
for $0 \le R \le L$, only the two states $|\,R\,\rangle$ and $|\,R-L\,\rangle$
compete for the minimum, while all other Wilson states have higher energy,
cf.~figure \ref{fig:ym11}. Since the Coulomb Hamiltonian is color diagonal
and field-independent for these states, we can complete the calculation of the
relevant matrix elements as follows:
\begin{equation}
\mathrm{tr}\,\langle\,z\,|\,e^{- T (H  + H_C)}\,|\,z\,\rangle =
\mathrm{tr}\,e^{- T H_C(z)}\cdot \langle \,z\,|\,e^{- T H}\,|\,z\,\rangle
=e^{- T V_C(z)}\cdot \sum_{n=0}^\infty e^{- T (E_n - E_0)}\,
\mathrm{tr}\,\left|\langle\,n\,|\,U(z)\,|\,0\,\rangle\right|^2 \,,
\label{pause}
\end{equation}
where $z=R$ or $z = R-L$ and $H_C(z) = V_C(z)\,\mathbbm{1}$.
In the last step, we have inserted a complete set
of eigenstates from the zero quark sector, since the Wilson state $|z\rangle$
made from transversal gauge fields is no longer in the $q\bar{q}$ sector and
expected to have overlap with the zero-charge ground state. In fact, the overlap
can be worked out explicitly with the result
\be
\tr\,\Big| \,\langle \,n\, | \,U(z)\,| \, 0\,\rangle\, \Big|^2 =
\frac{2}{\pi^2} \lk
\frac{z}{L} \rk^2 \lk 1 - (-)^n \cos \frac{z}{L} \pi \rk \lk \frac{1}{\lk
\frac{z}{L} \rk^2 - n^2} - \frac{1}{\lk \frac{z}{L} \rk^2 - (n + 2)^2} \rk^2 \,.
\label{G56}
\ee
The final sum over energy eigenvalues $n$ in eq.~(\ref{pause}) can only be
performed numerically. At large spacetime volumes $ L T \gg g^{-2}$, the ground
state $n=0$ dominates the sum, and all remaining contributions are
exponentially suppressed. The relevant matrix element
$|\langle\,0\,|\,R\,\rangle|^2$ is then independent of $T$ and approaches
unity for large space extensions $L \gg R$. This entails that
\[
- \lim_{T \to \infty}T^{-1}\,\ln\, \tr\,\langle\, z \,|\,e^{- T(H-E_0)}\,|\,z\,\rangle
= V_C(z)\,,\qquad\qquad z \in \{ R\,, R - L \}\,.
\]
The prescription (\ref{truepot11}) (with only $m=0$ and $m=-1$ competing
for the minimum if $|R| < L$) yields the final result
\begin{equation}
V(R) = \min_{m \in \{0,-1\}}\,V_C(R + mL) =
 \min_{m \in \{0,-1\}}\,\left[\frac{3}{8} g^2 (R + m L) \,\left(1 -
\frac{1}{3} \frac{R + m L}{L}\right)\right]\,.
\end{equation}

\bigskip\noindent
This function is plotted in the right panel of figure \ref{fig:ym11}. It is
symmetric around $R = L/2$ and thus invariant under the replacement $R \to L-R$.
The previous construction also entails that $V(R)$ has to be extended periodically
outside the range $0 \le R \le L$, although we have not performed that calculation
explicitly. The level crossing at $R = L/2$ is clearly visible, when the
two Wilson states from figure \ref{fig:wilson11} exchange their roles.
The effective string tension is
\begin{equation}
\sigma_{1+1} \equiv \left.\frac{d V(R)}{dR}\right|_{R/L \to 0} =
\frac{3}{8}\,g^2
\end{equation}
and this agrees with eq.~(\ref{sigma11}) and calculations using path integrals 
\cite{Witten:1991we,Blau:1991mp} 
or explicit quark wave functions \cite{Engelhardt:1995qm}. In the infinite 
volume limit $L \to \infty$, the potential is exactly linear, since all Wilson 
states other than $|\,R\,\rangle$ have energy of order $\mathscr{O}(L)$ and thus
decouple from the spectrum.

\no
\section{Yang-Mills theory in $(3+1)$ dimensions}
\label{sec:YM31}
\no
Let us finally address the most complex example, viz.~non-Abelian Yang-Mills
theory in $(3+1)$ dimensions. We employ again the first method based on the
standard gauge fixed Hamiltonian eq.~(\ref{Hfix}),
\begin{equation}
\langle W \rangle = \mathrm{tr}\,\langle\,\vek{R}^\perp\,|\,e^{- T (H-E_0)}\,|\,\vek{R}\,\rangle
= \int \mathscr{D} A^\perp \mathscr{J}[A^\perp] \,\Psi_R[\vek{A}^\perp]\,
e^{- T (H-E_0)}\,\Psi_R[\vek{A}^\perp]\,.
\label{984}
\end{equation}
Here, $\mathscr{J}$ is the Faddeev-Popov determinant and
$
\label{991}
\Psi_{\kvR}[\vek{A}^\perp] = \langle\,\vek{A}^\perp\,|\, U(\vek{R})\,|\,0\,\rangle
$
is the wave function of the Wilson state $|\,\vR\, \rangle$ in Coulomb gauge.

To proceed, we could take a model for the true Yang-Mills ground state
in Coulomb gauge, $|\,0\,\rangle$ and compute $\Psi_R[\vek{A}^\perp]$ by
application of the Wilson line. Below, we will instead apply an approximation
where only the vev.~of the Hamiltonian (rather than
the full evolution operator) is required. In such cases, it is expedient to
perform the same unitary transformation eq.~(\ref{T1}) which transfers the
Wilson line in the Hamiltonian, this time however with the gauge fixed version
eq.~(\ref{Hfix}) of the initial Hamiltonian. Retracing the steps that led
from eq.~(\ref{G8}) to eq.~(\ref{W10}) with the parallel transporter
$U(\vek{R})$ now depending on $\vek{A}^\perp$ only, we find  that a transversal
electric field $\Vek{\epsilon}^\perp$ is induced
which shifts the momentum operator in eq.~(\ref{Hfix}). Moreover, the
transformation of the Coulomb term eq.~(\ref{HC}) replaces the full charge
$(\rho + \rho_{\rm dyn})$ with $(\widetilde{\rho} + \rho_{\rm dyn} + \rho_{\rm ind})$,
where $\widetilde{\rho}$ is the rotated external charge eq.~(\ref{rhotilde}),
$\rho_{\rm dyn} \equiv - [\,\hat{\vek{A}}^\perp\,,\,\Vek{\Pi}^\perp\,]$ and
$\rho_{\rm ind} \equiv - [\,\hat{\vek{A}}^\perp\,,\,\Vek{\epsilon}^\perp\,]$.
The effective gauge fixed Hamiltonian is therefore
\begin{equation}
\widetilde{H}^{\rm fix} = H_{YM}^{\rm fix} + \tilde{V}_C + \Delta H\,,
\label{Hperpfix}
\end{equation}
where the first piece is just the usual gauge-fixed Hamiltonian in the absence
of external charges,
\begin{eqnarray}
H_{\rm YM}^{\rm fix} &=& \frac{g^2}{2}\,\int d^3x\, \mathscr{J}^{-1}\,
\Vek{\Pi}^\perp_a(\vek{x})\,\mathscr{J}\,\Vek{\Pi}^\perp_a(\vek{x}) +
\frac{1}{2 g^2}\,\int d^3x\,\big[\,\vek{B}^a(\vek{x})\,\big]^2 +
\nonumber \\[2mm]
&&\qquad\qquad{}+
\frac{g^2}{2}\,\int d^3(x,y)\,\mathscr{J}^{-1}\,\rho^a_{\rm dyn}(\vek{x})\,
\mathscr{J}\cdot F^{ab}(\vek{x},\vek{y})\,\rho_{\rm dyn}^b(\vek{y})\,\,.
\end{eqnarray}
The two additional pieces comprise a non-Abelian Coulomb potential
\begin{equation}
\widetilde{V}_C = \frac{g^2}{2}\int d^3 (x,y)\,\widetilde{\rho}^a(\vek{x})\,
F^{ab}(\vek{x},\vek{y})\,\widetilde{\rho}^b(\vek{y})
\end{equation}
and a new contribution which contains the coupling of the external charges
$\widetilde{\rho}$ to the dynamical charge of the gluon, as well as the
terms induced by the electric field due to the Wilson line,
\begin{eqnarray}
\Delta H &=& \frac{g^2}{2} \int d^3 x\,\Big\{ \mathscr{J}^{-1}\,\Vek{\Pi}^\perp(\vek{x})\,
\mathscr{J}\,\Vek{\epsilon}^\perp(\vek{x}) + \Vek{\epsilon}^\perp(\vek{x})\,
\Vek{\Pi}^\perp(\vek{x}) + \big[\,\Vek{\epsilon}^\perp(\vek{x})\,\big]^2\Big\} +
\nonumber \\[2mm]
&&{} + \frac{g^2}{2}\,\int d^3(x,y)\,\Bigg[ \mathscr{J}^{-1}\,\rho_{\rm dyn}
\mathscr{J}\,F\,(\widetilde{\rho} + \rho_{\rm ind}) +
(\widetilde{\rho} + \rho_{\rm ind})\,F\,\rho_{\rm dyn} +
\nonumber \\[2mm]
&&\qquad\qquad\qquad\qquad + \rho_{\rm ind}\,F\,\rho_{\rm ind} +
\rho_{\rm ind}\,F\,\widetilde{\rho}
+ \widetilde{\rho}\,F\,\rho_{\rm ind} \Bigg]
\nonumber
\end{eqnarray}
While a full calculation of $\langle W \rangle$ cannot be done with a Hamiltonian as
complicated as eq.~(\ref{Hperpfix}), it naturally lends itself to approximations
based on models for the (uncharged) vacuum wave functional, since the whole
effect of the Wilson line is already incorporated in the dynamics, i.e.~in the 
transformed Hamiltonian $\widetilde{H}$.
In the following, we will sketch such an approximation based on a Gaussian
model for the vacuum wave functional, as obtained recently by variational
calculations \cite{Feuchter:2004mk,Feuchter:2004gb,Reinhardt:2004mm,%
Epple:2006hv,Epple:2007ut,Reinhardt:2008ij,Schleifenbaum:2006bq,%
Reinhardt:2007wh}.

In a first step, we use Jensen's inequality to obtain a lower bound for the
Wilson loop (and thus an upper bound for the Wilson potential),
\[
\langle\, W\, \rangle = \left\langle\,0\,\left|\,\tr\,e^{- T(\widetilde{H}^{\rm fix} - E_0)}\,
\right|\,0\,\right\rangle \ge \tr\,\exp\Big[ - \left\langle\,0\,\left|\,
T \,(\widetilde{H}^{\rm fix} - E_0)\,\right|\,0\,\right\rangle\Big]\,.
\]
The advantage of the rhs is that only the vev of the Hamiltonian is required.
In particular, the eigenvalue equation
$H_{\rm YM}^{\rm fix}\,|0\,\rangle = E_0\,|\,0\,\rangle$ entails that the
YM Hamiltonian drops out and we are left with
\begin{equation}
\langle\,W\,\rangle \ge \tr\,\exp\left[ - \left\langle\,0\,\left|\,
T \,(\widetilde{V}_C + \Delta H)\,\right|\,0\,\right\rangle\right]\,.
\label{Wjensen}
\end{equation}
To compute the exponent on the rhs of this equation, we write out 
$\Delta H$ explicitly,
\[
\langle\,0\,|\, \widetilde{V}_C + \Delta H\,|\,0\,\rangle = \langle\,0\,|\,
\widetilde{V}_C\,|\,0\,\rangle +
\Big\langle\,0\,\Big|\, \frac{g^2}{2}\int \left(\Vek{\epsilon}^\perp\Vek{\Pi}^\perp +
\Vek{\Pi}^\perp\Vek{\epsilon}^\perp
 + \mathscr{J}^{-1}\,[\,\Vek{\Pi}^\perp\,,\,\mathscr{J}\,]\,\Vek{\epsilon}^\perp
 + \big[\Vek{\epsilon}^\perp\big]^2 + \overline{\Delta H}\right)\,\Big|\,0\,\Big\rangle\,.
\]
(The explicit form of the remainder $\overline{\Delta H}$ will be given
below.) Next we neglect the commutators
\begin{equation}
[\,\Vek{\Pi}^\perp\,,\,\mathscr{J}\,] \approx 0\,,\qquad\quad
[\,U^\perp(\vek{R})\,,\,T^a\,] \approx 0
\label{approx1}
\end{equation}
where $U^\perp(\vek{R})$ is the Wilson line $\vek{0} \to \vek{R}$ built from
the transversal vector field $\vek{A}^\perp$.
The first equation means that the gauge field dependence of the Faddeev--Popov
determinant is neglected, $\mathscr{J} = \mbox{const}$, while the second equation
discards the matrix structure of the non-Abelian parallel transporter,
$U^\perp \sim \mathbbm{1}$. (This is stronger than just neglecting path ordering).
As a consequence, the external charge and the induced
electric field also become independent of the gauge field,
\begin{eqnarray}
\widetilde{\rho}^a(\vek{x}) &\to& i T^a\,\big[ \delta(\vek{x}) -
\delta(\vek{x}-\vek{R}) \big]
\nonumber\\[2mm]
\big[\epsilon^\perp\big]^a_i(\vek{x}) &\to&
i T^a\int dy_k\,t_{ik}(\vek{x})\,\delta(\vek{x}-\vek{y}) \,.
\label{approx2}
\end{eqnarray}
The relevant matrix element in the exponent of eq.~(\ref{Wjensen}) now reads
\begin{equation}
\langle\,0\,|\, \widetilde{V}_C + \Delta H\,|\,0\,\rangle \approx
\frac{g^2}{2} \int \widetilde{\rho}\,\overline{F}\,\widetilde{\rho}
+ \frac{g^2}{2}\int d^3x\,\big[ \Vek{\epsilon}^\perp_a(\vek{x})\big]^2 +
\langle \,0\,|\,\overline{\Delta H}\,|\,0\,\rangle
\label{intermediate}
\end{equation}
where we used global color and translation invariance to infer the structure
\begin{equation}
\langle\,0\,|\,\Vek{\Pi}^{\perp}_a(\vek{x})\,|\,0\,\rangle = 0\,,\qquad\qquad
\langle \,0\,|\,F^{ab}(\vek{x},\vek{y})\,|\,0\,\rangle = \delta^{ab}\,
\overline{F}(\vek{x}-\vek{y})\,.
\label{Fvev}
\end{equation}
The first term on the rhs of eq.~(\ref{intermediate}) is the 
non-Abelian Coulomb interaction, i.e.~the remaining terms describe, 
within our approximation scheme, the difference between the (non-Abelian)
Coulomb potential and the Wilson potential. We will now investigate these 
corrections:

\medskip\noindent
The third term on the rhs of eq.~(\ref{intermediate}) involves the coupling
of the various charges via the Coulomb kernel,
\be
\label{12-91}
\overline{\Delta H} = \frac{g^2}{2} \int \Big[\mathscr{J}^{- 1}
\rho_{\rm dyn} \mathscr{J} F \lk \tilde{\rho} + \rho_{\rm ind} \rk +
\lk \tilde{\rho}  + \rho_{\rm ind} \rk \,F\, \rho_{\rm dyn} +
\rho_{\rm ind} \,F\, \rho_{\rm ind} + \rho_{\rm ind} \,F\, \tilde{\rho} +
\tilde{\rho}\, F\, \rho_{\rm ind}\Big] \hk .
\ee
Its vev is still too complex for a direct evaluation. To further simplify it,
we make yet another approximation and assume that the Coulomb kernel, $F$,
inside charge interactions can be replaced by its vev eq.~(\ref{Fvev}). In
an obvious notation, this means
\begin{equation}
\langle \,0\,|\,Q^a\,F^{ab}\,Q'^b\,|\,0\rangle \approx
\langle\,0\,|\, Q^a\,Q'^b\,|\,0\,\rangle\cdot \langle\,0\,|\,
F^{ab}\,|\,0\,\rangle =
\langle \,0\,|\, Q^a\,Q'^a\,|\,0\,\rangle \cdot \overline{F}
\label{approx3}
\end{equation}
for any two operators $Q,Q'$. We can now use the previous approximation
eq.~(\ref{approx1}) and consequently eq.~(\ref{approx2}) to put
$\langle \,0\,|\,\overline{\Delta H}\,|\,0\rangle\,$ in the form
\[
\frac{g^2}{2} \int \Big[\,
\langle \rho_{\rm dyn}\rangle \,\overline{F}\,\widetilde{\rho} +
\langle \rho_{\rm dyn}\,\rho_{\rm ind}\rangle\,\overline{F} +
\widetilde{\rho}\,\overline{F}\,\langle\rho_{\rm dyn}\rangle +
\langle \rho_{\rm ind}\,\rho_{\rm dyn}\rangle\,\overline{F} +
\langle \rho_{\rm ind}\,\rho_{\rm ind}\rangle\,\overline{F} +
\langle \rho_{\rm ind}\rangle\,\overline{F}\,\widetilde{\rho} +
\widetilde{\rho}\,\overline{F}\,\langle \rho_{\rm ind}\rangle\,\Big]\,.
\]
Since Coulomb gauge does not single out a color or space direction, global
color and translation invariance tells us that
\[
\langle\,0\,|\,\vek{A}^\perp_a(\vek{x})\,|\,0\,\rangle = 0\,,\qquad\qquad
\langle\,0\,|\,\vek{A}^\perp_a(\vek{x})\,\Vek{\Pi}^\perp_b(\vek{x})\,|\,0\,\rangle
\sim \delta_{ab}
\]
so that $\langle\rho_{\rm dyn}\rangle = \langle \rho_{\rm ind}\rangle = 0$.
Furthermore, we have the charge correlator
\[
\langle\,\rho_{\rm dyn}\,\rho_{\rm ind}\,\rangle =
\langle\,\rho_{\rm ind}\,\rho_{\rm dyn}\,\rangle =
\epsilon_k^{\perp\,a}(\vek{x})\,\langle\,0\,|\,f^{abe}\,f^{edc}\,
\vek{A}_k^{\perp\,b}(\vek{x})\,\vek{A}_i^{\perp\,c}(\vek{x})\,\Vek{\Pi}^{\perp\,d}_i(\vek{x})
\,|\,0\,\rangle\,.
\]
(For the first equality, we have to commute $\Vek{\Pi}$ to the right and use
$\langle\vek{A}\rangle = 0$.) The vev is a single component of a color and
space vector, and again global color and (spatial) rotation invariance tells us that
it has to vanish. We are thus left with only a single term
\[
\langle \,0\,|\,\overline{\Delta H}\,|\,0\,\rangle =
\frac{g^2}{2}\,f^{abc}\,f^{ade}\int d^3(x,y)\,\epsilon_i^{\perp\,b}(\vek{x})\,
\epsilon_k^{\perp\,d}(\vek{x})\,\overline{F}(\vek{x}-\vek{y})\cdot
\langle\,0\,|\, A_i^{\perp\,c}(\vek{x})\,A_k^{\perp\,e}(\vek{y})\,|\,0\,\rangle\,.
\]
Using global color and translation invariance one last time, the gluon propagator
must take the form
\begin{equation}
\langle\,0\,|\,A_i^{\perp\,c}(\vek{x})\,A_k^{\perp\,e}(\vek{y})\,|\,0\,\rangle =
\delta^{ce}\,t_{ik}(\vek{x})\,D(|\vek{x}-\vek{y}|)
\label{gluonprop}
\end{equation}
with an unknown scalar function $D(r)$. Thus eventually,
with $f^{abc}\,f^{adc} = N_C\,\delta^{bd}$,
\begin{equation}
\langle \,0\,|\,\overline{\Delta H}\,|\,0\,\rangle =
\frac{g^2}{2}\,N_C\int \Vek{\epsilon}^\perp_a(\vek{x}) \,\overline{F}(\vek{x}-\vek{y})
\,D(\vek{x}-\vek{y})\,\Vek{\epsilon}^\perp_a(\vek{y})\,.
\label{DeltaHfinal}
\end{equation}
We observe that this expression can be combined with the second term in
eq.~(\ref{intermediate}). 

\medskip\noindent
The first term in eq.~(\ref{intermediate}) is the non-Abelian Coulomb
interaction. It can easily be computed within our approximation scheme, since the
simplified form eq.~(\ref{approx2}) for the external charge implies
\begin{equation}
\langle\,0\,|\,\widetilde{V}_C\,|\,0\,\rangle =
\frac{g^2}{2}\int \widetilde{\rho}\,\overline{F}\,\widetilde{\rho} =
- \frac{g^2}{2}\,C_2 \Big[\, \overline{F}(R) - \overline{F}(0)\,\Big]
\end{equation}
with the quadratic Casimir $C_2(N_c) = - \sum_a\,T^a\,T^a$.
Putting everything together, we arrive at
\begin{eqnarray}
\langle \,W\,\rangle &\approx& \exp\Bigg\{ - T\,\frac{g^2}{2}\,N_C\int d^3(x,y)\,
\Vek{\epsilon}_a^\perp(\vek{x})\,\Big[ \,N_C^{-1}\,\delta(\vek{x}-\vek{y}) +
\overline{F}(\vek{x}-\vek{y})\,D(\vek{x}-\vek{y})\Big]\,\Vek{\epsilon}_a^\perp(\vek{y})
\Bigg\} \times
\nonumber \\[2mm]
&&{} \times\exp\Bigg\{ - T \,C_2\,g^2 \,\Big[ \overline{F}(R) - \overline{F}(0)\,\Big]\,\Bigg\}\,.
\label{final31}
\end{eqnarray}
To estimate the contribution from the first line, we require some dynamical input on the gluon
propagator $D$ and the Coulomb kernel $\overline{F}$. (Within our approximation, this is the
only place where the actual form of the vacuum wave functional enters.) The exact analytical
expressions are unknown, but we have reliable data from recent variational calculations 
\cite{Feuchter:2004mk,Feuchter:2004gb,Reinhardt:2004mm,Epple:2006hv,
Epple:2007ut,Reinhardt:2008ij,Schleifenbaum:2006bq,Reinhardt:2007wh},
which are also in decent agreement with high precision lattice measurements 
\cite{Quandt:2007qd,Burgio:2008jr}. The latter give
\begin{eqnarray}
D(k) &=& \frac{1}{2 \,\omega(k)}\,,\qquad\qquad \omega(k) \approx 
\sqrt{k^2 + M^4 / k^2} \nonumber \\[2mm]
\overline{F}(k) &\simeq& 8 \pi\,\frac{\sigma_C}{k^4}\,,\qquad\qquad k \to 0\,,
\end{eqnarray}
where $M \approx 880\,\mathrm{MeV}$ is the Gribov mass and $\sigma_C$ is 
the Coulomb string tension, which is about $2\ldots3$ times the Wilson 
string tension.

With this input, it is shown in appendix \ref{appB} that the integral in the first line of eq.~(\ref{final31})
yields a Coulomb type of potential which decays as $1/R$ at large distances and is thus subleading
as compared to the second line (which gives a linearly rising potential). Thus,
\begin{equation}
\langle\,W\,\rangle \approx \mathrm{const}\cdot\exp\Big\{\,- T \,C_2 \,g^2 \,\overline{F}(R)
+ \mathrm{subleading}\Big\}
\label{veryfinal31}
\end{equation}
and the Wilson string tension equals the Coulomb string tension.

\bigskip\noindent
This simple result should, however, be taken with some caution: In deriving it, we had to
make a number of approximations,
\begin{enumerate}
\item Jensen's inequality eq.~(\ref{Wjensen})
\item the commutator approximations eq.~(\ref{approx1})
\item the factorization eq.~(\ref{approx3})
\item the variational solutions for the Green functions $D(r)$ and $\overline{F}(\vek{x})\,$
\end{enumerate}
The corresponding errors are not all under good control: While lattice calculations indicate that
the variational solutions in item \#4 are close to the exact results, Jensen's inequality \#1 entails
that our potential is, a priori, only an upper bound to the true Wilson potential, and it is
hard to predict the quality of that bound. Even more severe is the factorization \#3,
since it causes most of the couplings between the various charges to vanish. In particular, we
have lost the coupling of the external charges at the end points of the Wilson lines with
the dynamical charges of the gluon, as described by the contribution
\be
\label{1161}
\frac{g^2}{2} \int \Big[ \mathscr{J}^{-1} \rho_{\rm dyn} \mathscr{J} F \,\widetilde{\rho} +
\widetilde{\rho} \,F \,\rho_{\rm dyn} \Big] \hk .
\ee
In perturbation theory, one can show explicitly that this term reduces the Coulomb potential
$\frac{1}{2}\,g^2\int \rho\, F \,\rho$. It is tempting to expect that eq.~(\ref{1161}) will screen
the Coulomb interaction when treated non-perturbatively, and that this effect provides the
major reduction of the Coulomb string tension to the Wilsonian one.
A detailed study of the screening of external color charges by dynamical gluons is subject
to ongoing research.
\bi

\section{Conclusions}
\label{conclusions}
In this paper, we have investigated the temporal Wilson loop in the 
Hamiltonian formulation of Yang-Mills theory. Using a unitary transformation
we have demonstrated that the effect of the Wilson lines is to introduce a 
new non-Abelian (field-dependent) electric field in the Hamiltonian. This 
formulation leads to the correct result in some simple cases where the exact 
gauge-invariant ground state is known. 

\medskip
Without the exact ground state at hand, we need to fix the gauge, for which
Coulomb gauge is a particularly useful choice: Since the only constraint 
on physical states is Gauss' law which can be resolved exactly, any
normalizable \emph{Ansatz} for the ground state wave functional is admissible and the 
formulation is ideal for variational approaches. We have shown how to treat 
the exactly solvable cases in Coulomb gauge, and applied these methods 
to the physically important case of Yang-Mills theory in $(3+1)$ dimensions
using the optimal Gaussian wave functional from recent variational approaches
as input. Unfortunately, the dynamics of the system was still too complex so that 
additional approximations were necessary to find an analytical result. 
Within these approximation, the Wilson string tension was found to match the 
Coulomb string tension, $\sigma \approx \sigma_c$. Finally, we have discussed 
the qualitative effect of the neglected couplings in the Hamiltonian, 
which are expected to suppress the Wilson string tension towards the 
value $\sigma / \sigma_c \simeq 0.3\,\ldots\,0.5$ favored by lattice simulations.
A closer investigation of those missing couplings is currently underway.

\begin{appendix}
\section{The Abelian Coulomb potential in spherical coordinates}
\label{appA}
\bi

\no
It is convenient to use spherical coordinates $\vx = (r, \hat{x}) , \vR = (R,
\hat{\vR})$ and express the $\delta$-function by
\be
\label{1746}
\delta (\vx - s \vR) = \frac{\delta (r - s R)}{r^2} \sli_{l, m} Y^*_{l m}
(\hat{x}) Y_{l m} (\hat{R}) \hk .
\ee
With this representation we find with (\ref{staticpot})
\bea
\label{1752}
\frac{g^2}{2} \int d^3 x\, \Vek{\epsilon}(\vek{x})^2 & = & \frac{g^2}{2} R^2 \il^1_0 d s
\il^1_0 d s' \il^\infty_0 d r r^2 \frac{\delta (r - s R)}{r^2} \frac{\delta (r
- s'R)}{r^2} \times \nonumber\\[2mm]
& &{}\times  \sli_{l m} \sli_{l' m'} \int d \Omega_x Y^*_{l m} (\hat{x}) Y_{l'm'}
(\hat{x}) Y_{lm} (\hat{R}) Y^*_{l' m'} (\hat{R}) \hk .
\eea
Performing the integrals over $r, s'$ and $\Omega_x$ by means of the
orthonormality of $Y_{l m} (\hat{x})$ we obtain
\be
\label{F5-X2}
\frac{g^2}{2} \int d^3 x\, \Vek{\epsilon}(\vek{x})^2 = \frac{g^2}{2} \frac{1}{R} \il^1_0
\frac{d s}{s^2} \,\delta^{(2)}_{\sphericalangle} (\hat{0}) \hk ,
\ee
where $\delta^{(2)}_{\sphericalangle}$ is defined by
\be
\label{1768}
\delta^{(2)}_{\sphericalangle} (\Omega - \Omega')  =  \sli_{l m} Y^*_{l m}
(\hat{\Omega}) Y_{l m} (\Omega')
= \delta (\cos \vartheta - \cos \vartheta') \delta (\varphi - \varphi') \hk
.
\ee
Note that $\delta^{(2)}_{\sphericalangle} (\hat{\Omega})$ is dimensionless.
From the final integral
\be
\label{1777}
\left. \il^1_0 \frac{d s}{s^2} = - \frac{1}{s} \right|^1_{s = 0} = - 1 + \infty
\ee
we ignore the divergent contribution from the lower integration limit, which
represents the \emph{self-energy} contribution of the two charges. 
The divergent factor $\delta^{(2)}_{\sphericalangle}(\hat{0})$ has a different 
origin: it is due to the infinitely thin Wilson line between the charges
and gives rise to the renormalization of the Wilson loop operator, 
cf.~eq.~(\ref{Wbare}). In the present context, this amounts to the 
renormalization of the bare charge,
\be
\label{1785}
g^2 \,\delta^{(2)}_{\sphericalangle}
(\hat{0}) = \frac{e^2}{4 \pi} \hk ,
\ee
so that the Coulomb potential from eq.~(\ref{F5-X2}) becomes eventually
\be
\label{1791}
V (R) = \frac{g^2}{2} \int d^3 x\, \Vek{\epsilon}(\vek{x})^2
= - \frac{e^2}{4 \pi R} \hk
.
\ee
Our result (\ref{F5-X2}) disagrees with the similar calculation 
in ref.~\cite{Gaete:1998vr}, where the divergent factor 
$\delta^{(2)}_\sphericalangle (\hat{0})$ is replaced by 
$1/(4 \pi)$, so that no renormalization seems necessary 
(except for the subtraction of the self-energy).

\smallskip
To relate the present calculation to our reasoning in section \ref{general}
of the main text, it should be noted that eq.~(\ref{1746}) implies
$s = r / R = |\vek{x}|/R$ and the lower bound $s=0$ in eq.~(\ref{1777})
should therefore be $s_{\rm min} = |\vek{x}|_{\rm min} / R$, with
$|\vek{x}|_{\rm min}$ a minimal distance of the two charges. If we 
put a cutoff $s_{\rm min} = a/R$ for the lower integration limit 
in eq.~(\ref{1777}), the (unrenormalized) potential becomes
\[
V(R) = - \frac{g^2}{2}\,\delta^{(2)}_{\sphericalangle} (\hat{0})\,
\left(\frac{1}{R}-\frac{1}{a}\right)\,.
\]
Clearly, the self-energy of the charges corresponds to the second term
in the parenthesis. However, even with a minimal distance cutoff $a > 0$, 
there is still a divergent prefactor coming from the infinitely thin 
Wilson line. To see this, recall that $\Vek{\epsilon}(\vek{x})$ has support 
on the Wilson line only, so that $\hat{\vek{x}} = \hat{\vek{R}}$
if the Wilson line is infinitely thin, and hence the argument of the angular 
delta-function $\delta^{(2)}_{\sphericalangle}$ in eq.~(\ref{F5-X2}) vanishes.
The divergent prefactor is 
thus well understood as a typical OPE divergence for the thin Wilson line 
which involves field operators $\vek{A}(\vek{x})$ at arbitrarily close 
positions. It must be absorbed in the multiplicative renormalization 
$Z_W$ of the Wilson loop operator,
cf.~the discussion following eq.~(\ref{Wbare}) in the main text.
\bi

\no
\section{Contribution of $\Vek{\epsilon}^\perp$ to the Coulomb potential}
\label{appB}
To estimate the role of $\Vek{\epsilon}^\perp$ we calculate the vacuum expectation
value of the induced Coulomb term
\be
\label{1806}
\Delta H^{(1)}_c = \frac{g^2}{2} \int \hat{\vA} \cdot \Vek{\epsilon}^\perp F
\hat{\vA} \cdot \Vek{\epsilon}^\perp \hk .
\ee
Using the approximation
\be
\label{1812}
\langle \hat{A} F \hat{A} \rangle \simeq \langle \hat{A} \hat{A} \rangle \langle
F \rangle \equiv \frac{\overline{F}}{2 \omega}
\ee
and ignoring the path ordering in $\Vek{\epsilon}^\perp (x)$, i.e.
$\widetilde{T}_a (x) \simeq T_a$, we obtain
\be
\label{F13X2}
\langle \Delta H^{(1)}_c \rangle = C_2\, \frac{g^2}{2}
\il^{\mbox{\boldmath$\scriptstyle R$\unboldmath}}_0
d x_i
 \il^{\mbox{\boldmath$\scriptstyle R$\unboldmath}}_0
  d y_j \, t_{i j} (x) \omega^{- 1} (x, y) \, \overline{F} (x, y) \hk ,
\ee
where $C_2\,\mathbbm{1} = - T_a\,T_a$ is the quadratic Casimir operator
in the color representation of the Wilson loop.
Putting the path along the $z$-axis from $0$ to $R$
\be
\label{1833}
x_i = \delta_{i 3} R s \hk , \qquad  \hk y_i = \delta_{i 3} R t
\ee
we find after Fourier transformation and using spherical coordinates in momentum
space
\bea
\label{1839}
\langle \Delta H^{(1)}_c \rangle & = & R^2 \lk \frac{1}{2 \pi} \rk^4
\il^\infty_0 \frac{d k k^2}{\omega (k)} \il^\infty_0 d p\, p^2 F (p) \il^1_0 d s
\il^1_0 d t \times \nonumber\\
& & {}\times \il^1_{- 1} d z \,(1 - z^2) e^{i z k R (s - t)} \il^1_{- 1}
d z'\, e^{i z' p R (s - t)}\,.
\eea
We are interested in the large $R$ behavior of this quantity.
Assuming the usual IR-behavior of the Green functions,
\be
\label{1848}
\omega (k) \sim \frac{1}{k} \hk , \qquad
\overline{F} (k) \sim \frac{1}{k^4}
\ee
as $k \to 0$, and furthermore introducing  dimensionless variables
\be
\label{1853}
\bar{k} = k R \hk , \qquad \bar{p} = p R\,,
\ee
on finds
\be
\langle \Delta H^{(1)}_c \rangle \sim \frac{1}{R} \hk  \qquad
\quad\mbox{as\quad}R \to \infty \hk .
\label{1872}
\ee
This quantity is therefore subleading at large distances and
hence irrelevant for the screening of the Coulomb string tension.

\end{appendix}

\bibliographystyle{apsrev4-1}
\bibliography{references}

\end{document}